\documentclass[prd,preprintnumbers,twocolumn,amsmath,nofootinbib,amssymb]{revtex4}
\usepackage{graphicx,color,dcolumn,booktabs,bm}
\usepackage{longtable,lscape}
\usepackage{txfonts}
\usepackage{overpic}
\usepackage{amssymb}
\usepackage{epstopdf}
\usepackage{indentfirst}
\usepackage{feynmf}   
\usepackage{slashed}  
\usepackage{cases}
\usepackage{color}
\usepackage{float}
\usepackage{multirow}
\usepackage{ulem}
\usepackage{graphicx,color,dcolumn,booktabs,bm}
\usepackage{epsfig,dsfont,amssymb,amsmath,amsfonts,amsbsy,mathrsfs}

\graphicspath{{Figures/}} %

\usepackage{hyperref}
\hypersetup{colorlinks,citecolor=blue,anchorcolor=red,menucolor=red, linkcolor=red,filecolor=red,runcolor=red,urlcolor=blue,frenchlinks=red}

\makeatletter
\@addtoreset{equation}{section}
\makeatother

\allowdisplaybreaks

\begin{document}
	
\title{\textbf{Investigating the M1 radiative decay behaviors and the magnetic moments of the predicted triple-charm molecular-type pentaquarks} }
	
\author{Bao-Jun Lai$^{1,2,3,5}$}
\email{laibj2023@lzu.edu.cn}
\author{Fu-Lai Wang$^{1,2,3,5}$}
\email{wangfl2016@lzu.edu.cn}
\author{Xiang Liu$^{1,2,3,4,5}$}
\email{xiangliu@lzu.edu.cn}
\affiliation{$^1$School of Physical Science and Technology, Lanzhou University, Lanzhou 730000, China\\
$^2$Lanzhou Center for Theoretical Physics, Key Laboratory of Theoretical Physics of Gansu Province, Lanzhou University, Lanzhou 730000, China\\
$^3$Key Laboratory of Quantum Theory and Applications of MoE, Lanzhou University,
Lanzhou 730000, China\\
$^4$MoE Frontiers Science Center for Rare Isotopes, Lanzhou University, Lanzhou 730000, China\\
$^5$Research Center for Hadron and CSR Physics, Lanzhou University and Institute of Modern Physics of CAS, Lanzhou 730000, China}
	
\begin{abstract}
In this work, we systematically study the electromagnetic properties including the M1 radiative decay widths and the magnetic moments of the isoscalar $\Xi_{c c} D^{(*)}$, $\Xi_{cc}D_{1}$, and $\Xi_{cc}D_{2}^{*}$ triple-charm molecular-type pentaquark candidates, where we adopt the constituent quark model and consider both the $S$-$D$ wave mixing effect and the coupled channel effect. Our numerical results suggest that the M1 radiative decay widths and the magnetic moments of the isoscalar $\Xi_{c c} D^{(*)}$, $\Xi_{cc}D_{1}$, and $\Xi_{cc}D_{2}^{*}$ triple-charm molecular-type pentaquark candidates can reflect their inner structures, and the study of the electromagnetic properties is the important step to construct the family of the triple-charm molecular-type pentaquarks. With the accumulation of the experimental data during the high-luminosity phase of LHC, we expect that the present work combined with the corresponding mass spectrum information can encourage the experimental colleagues at LHCb to focus on the isoscalar $\Xi_{c c} D^{(*)}$, $\Xi_{cc}D_{1}$, and $\Xi_{cc}D_{2}^{*}$ triple-charm molecular-type pentaquark candidates. 
\end{abstract}

\maketitle

\section{Introduction}\label{sec1}

With the observation of more and more new hadronic states \cite{Liu:2013waa,Hosaka:2016pey,Chen:2016qju,Richard:2016eis,Lebed:2016hpi,Brambilla:2019esw,Liu:2019zoy,Chen:2022asf,Olsen:2017bmm,Guo:2017jvc,Meng:2022ozq}, the study of hadron spectroscopy has entered a new era, representing the high-precision frontier of particle physics. A key objective of these investigations is to provide crucial information to deepen our understanding of the non-perturbative behavior of the strong interaction. It is still full of challenges and opportunities to the whole community.
	
In the past decade, big progress has been made in exploring the hidden-charm molecular-type pentaquarks, thanks to the joint efforts of both theoretical \cite{Li:2014gra,Karliner:2015ina,Wu:2010jy,Wang:2011rga,Yang:2011wz,Wu:2012md,Chen:2015loa} and experimental \cite{Aaij:2015tga,Aaij:2019vzc,LHCb:2020jpq,LHCb:2022ogu} sides. Staying in the new stage, more predictions around molecular-type pentaquarks have been given \cite{Liu:2013waa,Hosaka:2016pey,Chen:2016qju,Richard:2016eis,Lebed:2016hpi,Brambilla:2019esw,Liu:2019zoy,Chen:2022asf,Olsen:2017bmm,Guo:2017jvc,Meng:2022ozq}. Among them, a typical example is the predicted triple-charm molecular-type pentaquarks \cite{Chen:2017jjn,Wang:2019aoc}, which is due to the interaction of double-charm baryon and charmed meson. Indeed, this study is also motivated by the observation of  the doubly charmed baryon $\Xi_{cc}(3620)^{++}$ reported by the LHCb Collaboration in 2017 \cite{LHCb:2017iph}. Although the mass spectrum of triple-charm molecular-type pentaquarks has been given \cite{Chen:2017jjn,Wang:2019aoc}, we still need to make more efforts to provide theoretical suggestions for searching for them.  

We can note a recent development in the study of the spectroscopy of molecular-type pentaquarks. For example, in Refs. \cite{Wang:2016dzu,Li:2021ryu,Ozdem:2021ugy,Xu:2020flp,Zhou:2022gra,Gao:2021hmv,Wang:2022nqs,Ozdem:2023htj,Ozdem:2022kei,Wang:2022tib,Wang:2023aob,Wang:2023ael,Guo:2023fih,Li:2024wxr,Li:2024jlq,Ozdem:2024jty,Ozdem:2024rqx}, the authors suggested that the electromagnetic properties of 
molecular-type pentaquarks should be paid more attention. One main reason is that the electromagnetic properties, such as the magnetic moments of the hadron, may reflect its inner structure. More importantly, their M1 radiative decays are accessible in experiments. 

Along this line, in this work we carry out the investigation of the electromagnetic properties of the predicted triple-charm molecular-type pentaquarks \cite{Chen:2017jjn,Wang:2019aoc}. We focus mainly on quantitative calculation of their M1 radiative decay behavior, since these physical quantities can be measured in experiments such as LHCb during the high-luminosity phase of LHC \cite{Bediaga:2018lhg}. Although it is a challenging task to directly measure their magnetic moments, this does not prevent us from carrying out the phenomenological study of this question. We hope that the present work, combined with earlier work \cite{Chen:2017jjn,Wang:2019aoc} on the corresponding mass spectrum, will provide more complete spectroscopic information on triple-charm molecular-type pentaquarks.  
	
The rest of this paper is organised as follows. In Sec. \ref{sec2}, we focus mainly on quantitative calculation of the M1 radiative decay widths of the isoscalar $\Xi_{c c} D^{(*)}$, $\Xi_{cc}D_{1}$, and $\Xi_{cc}D_{2}^{*}$ triple-charm molecular-type pentaquark candidates. In Sec. \ref{sec3}, we discuss the magnetic moments of the isoscalar $\Xi_{c c} D^{(*)}$, $\Xi_{cc}D_{1}$, and $\Xi_{cc}D_{2}^{*}$ triple-charm molecular-type pentaquark candidates. Finally, we give the discussion and conclusion in Sec. \ref{sec4}.

\section{The M1 radiative decay widths of the triple-charm molecular-type pentaquarks}\label{sec2}

In Refs. \cite{Chen:2017jjn,Wang:2019aoc}, the mass spectrum of the $\Xi_{c c} D^{(*)}$, $\Xi_{cc}D_{1}$, and $\Xi_{cc}D_{2}^{*}$ triple-charm molecular-type pentaquarks was predicted based on the one-boson-exchange model, taking into account the $S$-$D$ wave mixing effect and the coupled channel effect. Usually, the loosely bound state with the cutoff value around 1 GeV can be considered as the most promising molecular candidate according to the experience of the deuteron studies \cite{Machleidt:1987hj,Epelbaum:2008ga,Esposito:2014rxa,Chen:2016qju,Tornqvist:1993ng,Tornqvist:1993vu,Wang:2019nwt,Chen:2017jjn}, the authors suggested that the $\Xi_{cc}D$ state with $I(J^P)=0(1/2^-)$, the $\Xi_{cc}D^{*}$ state with $I(J^P)=0(3/2^-)$, the $\Xi_{cc}D_{1}$ states with $I(J^P)=0(1/2^+,3/2^+)$, and the $\Xi_{cc}D_{2}^{*}$ states with $I(J^P)=0(3/2^+,5/2^+)$ can be recommended as the most promising candidates of the triple-charm molecular-type pentaquarks \cite{Chen:2017jjn,Wang:2019aoc}. At present, our understanding of the properties of the triple-charm molecular-type pentaquark candidates is not sufficient. Thus, it is necessary to provide more comprehensive suggestions to encourage the experimental colleagues to explore the triple-charm molecular-type pentaquarks.

The study of the electromagnetic properties, such as the radiative decay width and the magnetic moment, can provide the essential insights for understanding the inner structures of the hadronic states, which is crucial for distinguishing the spin-parity quantum numbers and the configurations of the hadronic states. Therefore, the study of the electromagnetic properties is the important step for the experimental construction of the hadron family. For example, the ratio $R_{\gamma\psi}= \frac{\mathcal{B}[X(3872) \rightarrow \gamma\psi(2S)]}{\mathcal{B}[X(3872) \rightarrow \gamma J/\psi]}$ is essential for understanding the inner structure of the charmonium-like state $X(3872)$ \cite{BaBar:2008flx,LHCb:2014jvf,Belle:2011wdj,BESIII:2020nbj}. In this study, we investigate the electromagnetic properties including the M1 radiative decay widths and the magnetic moments of the isoscalar $\Xi_{c c} D^{(*)}$, $\Xi_{cc}D_{1}$, and $\Xi_{cc}D_{2}^{*}$ triple-charm molecular-type pentaquark candidates based on their mass spectra and spatial wave functions \cite{Chen:2017jjn,Wang:2019aoc}. In our concrete calculations, we take the constituent quark model, which is a reliable tool for discussing the electromagnetic properties of the hadrons in the past decades \cite{Liu:2003ab,Huang:2004tn,Zhu:2004xa,Haghpayma:2006hu,Wang:2016dzu,Deng:2021gnb,Gao:2021hmv,Zhou:2022gra,Wang:2022tib,Li:2021ryu,Schlumpf:1992vq,Schlumpf:1993rm,Cheng:1997kr,Ha:1998gf,Ramalho:2009gk,Girdhar:2015gsa,Menapara:2022ksj,Mutuk:2021epz,Menapara:2021vug,Menapara:2021dzi,Gandhi:2018lez,Dahiya:2018ahb,Kaur:2016kan,Thakkar:2016sog,Shah:2016vmd,Dhir:2013nka,Sharma:2012jqz,Majethiya:2011ry,Sharma:2010vv,Dhir:2009ax,Simonis:2018rld,Ghalenovi:2014swa,Kumar:2005ei,Rahmani:2020pol,Hazra:2021lpa,Gandhi:2019bju,Majethiya:2009vx,Shah:2016nxi,Shah:2018bnr,Ghalenovi:2018fxh,Wang:2022nqs,Mohan:2022sxm,An:2022qpt,Kakadiya:2022pin,Wu:2022gie,Wang:2023bek,Wang:2023aob}, particularly successfully reproducing the experimental data of the magnetic moments of the decuplet and octet baryons \cite{Schlumpf:1993rm,Ramalho:2009gk}.

In this section, we will study the M1 radiative decay widths of the isoscalar $\Xi_{c c} D^{(*)}$, $\Xi_{cc}D_{1}$, and $\Xi_{cc}D_{2}^{*}$ triple-charm molecular-type pentaquark candidates. In the calculations, the M1 radiative decay width between the hadronic states $\Gamma_{H \to H^{\prime}\gamma}$ can be related to the corresponding transition magnetic moment $\mu_{H \to H^{\prime}}$ \cite{Dey:1994qi,Simonis:2018rld,Gandhi:2019bju,Hazra:2021lpa,Li:2021ryu,Zhou:2022gra,Wang:2022tib,Rahmani:2020pol,Menapara:2022ksj,Menapara:2021dzi,Gandhi:2018lez,Majethiya:2011ry,Majethiya:2009vx,Shah:2016nxi,Ghalenovi:2018fxh,Wang:2022nqs,Mohan:2022sxm,An:2022qpt,Kakadiya:2022pin,Wang:2023bek,Wang:2023aob}, which can be expressed as \cite{Wang:2022nqs,Wang:2023bek,Wang:2023aob,Wang:2023ael}
\begin{eqnarray}\label{width}
	\Gamma_{H \to H^{\prime}\gamma}= \frac{\alpha_{\rm {EM}}}{2J_{H}+1}\frac{4k^{3}}{e^{2}}\frac{\sum\limits_{J_{H^{\prime}z},J_{Hz}}\left(\begin{array}{ccc} J_{H^{\prime}}&1&J_{H}\\-J_{H^{\prime}z}&0&J_{Hz}\end{array}\right)^2}{\left(\begin{array}{ccc} J_{H^{\prime}}&1&J_{H}\\-J_{z}&0&J_{z}\end{array}\right)^2}\left|\mu_{H \to H^{\prime}}\right|^2.\nonumber\\
\end{eqnarray}
In the above expression, $\alpha_{\rm EM}=1/137$ is the electromagnetic fine structure constant, $k=(m_{H}^{2}-m_{H^{\prime}}^{2})/(2m_{H})$ is the momentum of the emitted photon, and the symbol $\left(\begin{array}{ccc} a&b&c\\d&e&f\end{array}\right)$ is the $3$-$j$ coefficient. 

Based on the above discussions, it is necessary to calculate the corresponding transition magnetic moment when discussing the M1 radiative decay width between the hadronic states. At present, there is a lack of the experimental data for the isoscalar $\Xi_{c c} D^{(*)}$, $\Xi_{cc}D_{1}$, and $\Xi_{cc}D_{2}^{*}$ triple-charm molecular-type pentaquark candidates \cite{ParticleDataGroup:2022pth}, but the transition magnetic moment and the M1 radiative decay width between the hadronic molecular states depend on the binding energies of the initial and final molecules \cite{Wang:2022nqs,Wang:2023bek,Wang:2023aob,Wang:2023ael}. For the isoscalar $\Xi_{c c} D^{(*)}$, $\Xi_{cc}D_{1}$, and $\Xi_{cc}D_{2}^{*}$ triple-charm molecular-type pentaquark candidates, their transition magnetic moments can be determined by the following equation \cite{Wang:2022nqs,Wang:2023bek,Wang:2023aob,Wang:2023ael}
\begin{eqnarray}
\mu_{H \to H^{\prime}}=\left\langle{J_{H^{\prime}},J_{z}\left|\sum_{j}\hat{\mu}_{zj}^{\rm spin}e^{-i {\bf k}\cdot{\bf r}_j}+\hat{\mu}_z^{\rm orbital}\right|J_{H},J_{z}}\right\rangle^{J_{z}=\mathrm{min}\{J_{H},J_{H^{'}}\}}\label{transitionmagnetic}
\end{eqnarray}
when taking the same binding energy for the initial and final triple-charm molecular-type pentaquark candidates. Here, the magnetic moment operators $\hat{\mu}_{zj}^{\rm spin}$ and $\hat{\mu}_z^{\rm orbital}$ can be expressed as \cite{Liu:2003ab,Huang:2004tn,Zhu:2004xa,Haghpayma:2006hu,Wang:2016dzu,Deng:2021gnb,Gao:2021hmv,Zhou:2022gra,Wang:2022tib,Li:2021ryu,Schlumpf:1992vq,Schlumpf:1993rm,Cheng:1997kr,Ha:1998gf,Ramalho:2009gk,Girdhar:2015gsa,Menapara:2022ksj,Mutuk:2021epz,Menapara:2021vug,Menapara:2021dzi,Gandhi:2018lez,Dahiya:2018ahb,Kaur:2016kan,Thakkar:2016sog,Shah:2016vmd,Dhir:2013nka,Sharma:2012jqz,Majethiya:2011ry,Sharma:2010vv,Dhir:2009ax,Simonis:2018rld,Ghalenovi:2014swa,Kumar:2005ei,Rahmani:2020pol,Hazra:2021lpa,Gandhi:2019bju,Majethiya:2009vx,Shah:2016nxi,Shah:2018bnr,Ghalenovi:2018fxh,Wang:2022nqs,Mohan:2022sxm,An:2022qpt,Kakadiya:2022pin,Wu:2022gie,Wang:2023bek,Wang:2023aob,Wang:2023ael}
\begin{eqnarray}
\hat{\mu}_{zj}^{\rm spin}&=&\frac{e_j}{2m_j}\hat{\sigma}_{zj},\\
\hat{\mu}_z^{\rm orbital}&=&\left(\frac{m_{m}}{m_{b}+m_{m}}\frac{e_b}{2m_b}+\frac{m_{b}}{m_{b}+m_{m}}\frac{e_m}{2m_m}\right)\hat{L}_z,
\end{eqnarray}
where $\hat{\sigma}_{zj}$, $e_j$, and $m_j$ represent the $z$-component of the Pauli operator, the charge, and the mass of the $j$-th constituent, respectively. The subscripts $b$ and $m$ represent the doubly-charmed baryon and the charmed meson, while $L_{z}$ stands for the $z$-component of the orbital angular momentum operator between the doubly-charmed baryons and the charmed mesons. The expression $e^{-i {\bf k}\cdot{\bf r}_j}$ represents the spatial wave function of the emitted photon. 

In this study, we also discuss the magnetic moments of the isoscalar $\Xi_{c c} D^{(*)}$, $\Xi_{cc}D_{1}$, and $\Xi_{cc}D_{2}^{*}$ triple-charm molecular-type pentaquark candidates based on their mass spectra and spatial wave functions \cite{Chen:2017jjn,Wang:2019aoc}. Similarly to the method used to estimate the transition magnetic moment between the hadrons $\mu_{H \to H^{\prime}}$, the magnetic moment of the hadron $\mu_{H}$ within the constituent quark model can be deduced by the following relation \cite{Liu:2003ab,Huang:2004tn,Zhu:2004xa,Haghpayma:2006hu,Wang:2016dzu,Deng:2021gnb,Gao:2021hmv,Zhou:2022gra,Wang:2022tib,Li:2021ryu,Schlumpf:1992vq,Schlumpf:1993rm,Cheng:1997kr,Ha:1998gf,Ramalho:2009gk,Girdhar:2015gsa,Menapara:2022ksj,Mutuk:2021epz,Menapara:2021vug,Menapara:2021dzi,Gandhi:2018lez,Dahiya:2018ahb,Kaur:2016kan,Thakkar:2016sog,Shah:2016vmd,Dhir:2013nka,Sharma:2012jqz,Majethiya:2011ry,Sharma:2010vv,Dhir:2009ax,Simonis:2018rld,Ghalenovi:2014swa,Kumar:2005ei,Rahmani:2020pol,Hazra:2021lpa,Gandhi:2019bju,Majethiya:2009vx,Shah:2016nxi,Shah:2018bnr,Ghalenovi:2018fxh,Wang:2022nqs,Mohan:2022sxm,An:2022qpt,Kakadiya:2022pin,Wu:2022gie,Wang:2023bek,Wang:2023aob,Wang:2023ael}
\begin{eqnarray}
\mu_{H}&=&\left\langle{J_{H},J_{H}\left|\sum_{j}\hat{\mu}_{zj}^{\rm spin}+\hat{\mu}_z^{\rm orbital}\right|J_{H},J_{H}}\right\rangle.\label{magnetic}
\end{eqnarray}

\begin{figure}[htbp]
  \includegraphics[width=0.49\textwidth]{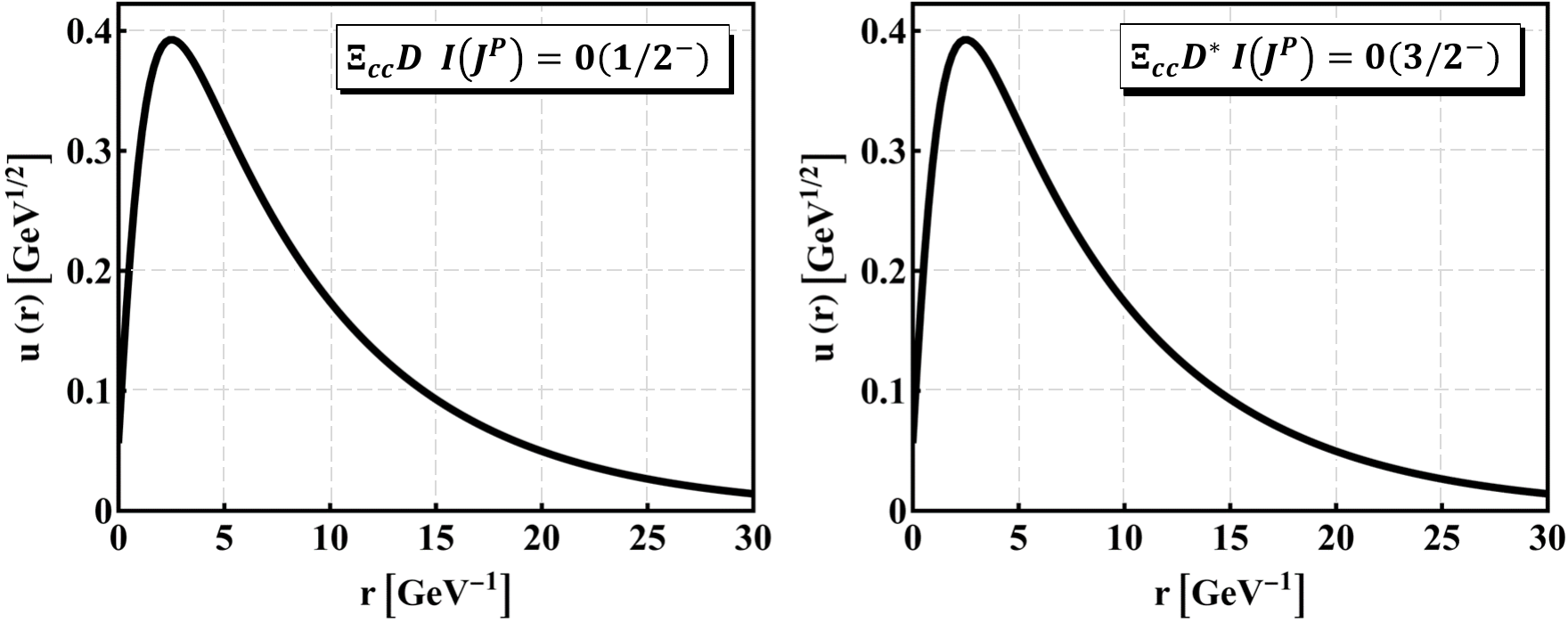}
  \caption{The spatial wave functions of the $\Xi_{cc}D$ state with $I(J^P)=0(1/2^-)$ and the $\Xi_{cc}D^{*}$ state with $I(J^P)=0(3/2^-)$ by solving the coupled channel Schr\"{o}dinger equation \cite{Chen:2017jjn}. Here, we take the binding energies of the $\Xi_{cc}D$ state with $I(J^P)=0(1/2^-)$ and the $\Xi_{cc}D^{*}$ state with $I(J^P)=0(3/2^-)$ as $-6.0~{\rm MeV}$.}\label{spatialwavefunctions}
\end{figure}

As shown in Eqs. (\ref{transitionmagnetic}) and (\ref{magnetic}), we need to construct the color, flavor, spin, and spatial wave functions of the hadronic states when calculating their transition magnetic moments and magnetic moments \cite{Wang:2022nqs,Wang:2023bek,Wang:2023aob,Wang:2023ael}. In this work, we utilize the simple harmonic oscillator wave function to approximate describe the spatial wave functions of the doubly-charmed baryons and the charmed mesons \cite{Wang:2022nqs,Wang:2023bek,Wang:2023aob,Wang:2023ael}. Explicitly, the simple harmonic oscillator wave function can be expressed as
\begin{eqnarray}
\phi_{n,l,m}(\beta,{\bf r})&=&\sqrt{\frac{2n!}{\Gamma(n+l+\frac{3}{2})}}L_{n}^{l+\frac{1}{2}}(\beta^2r^2)\beta^{l+\frac{3}{2}}{\mathrm e}^{-\frac{\beta^2r^2}{2}}r^l Y_{l m}(\Omega_{\bf r}).\nonumber\\
\end{eqnarray}
In the above formula, the symbols $n$, $l$, and $m$ are the radial, the orbital, and the magnetic quantum numbers of the discussed hadron, respectively. $L_{n}^{l+\frac{1}{2}}(x)$ and $Y_{l m}(\Omega_{\bf r})$ are the associated Laguerre polynomial and the spherical harmonic function, respectively. $\beta$ is a parameter in the simple harmonic oscillator wave function, and we take $\beta_{\Xi_{cc}\rho}=0.454~{\rm GeV}$, $\beta_{\Xi_{cc}\lambda}=0.427~{\rm GeV}$ \cite{Yu:2022lel}\footnote{For the $\Xi_{cc}$ baryon, the two charmed quarks are treated as the diquark, the $\rho$-mode corresponds to the excitation of the diquark, and the $\lambda$-mode corresponds to the diquark-quark excitation \cite{Yu:2022lel}.}, $\beta_{D}=0.601~{\rm GeV}$, $\beta_{D^{*}}=0.516~{\rm GeV}$, $\beta_{c \bar q|1^1P_{1}\rangle}=0.475~{\rm GeV}$, $\beta_{c \bar q|1^3P_{1}\rangle}=0.482~{\rm GeV}$, and $\beta_{c \bar q|1^3P_{2}\rangle}=0.437~{\rm GeV}$ \cite{Godfrey:2015dva} in the following numerical analysis. 
For the isoscalar $\Xi_{cc} D^{(*)}$, $\Xi_{cc}D_{1}$, and $\Xi_{cc}D_{2}^{*}$ triple-charm molecular-type pentaquark candidates, we take the precise spatial wave functions between two hadrons by solving the coupled channel Schr\"{o}dinger equation in our concrete calculations. Specifically, the spatial wave functions of the isoscalar $\Xi_{cc} D$ and $\Xi_{cc} D^{*}$ molecular states can be obtained from the quantitative study of their mass spectrum in Ref. \cite{Chen:2017jjn}, and the spatial wave functions of the $\Xi_{cc}D_{1}$ and $\Xi_{cc}D_{2}^{*}$ molecular states can be obtained based on the quantitative study of their mass spectrum in Ref. \cite{Wang:2019aoc}. For example, we present the obtained spatial wave functions of the $\Xi_{cc}D$ state with $I(J^P)=0(1/2^-)$ and the $\Xi_{cc}D^{*}$ state with $I(J^P)=0(3/2^-)$ by solving the coupled channel Schr\"{o}dinger equation in Fig. \ref{spatialwavefunctions}.

In the specific calculations, we will expand the spatial wave function of the emitted photon $e^{-i{\bf k}\cdot{\bf r}_j}$ as the following form \cite{Khersonskii:1988krb}
\begin{eqnarray}
e^{-i{\bf k}\cdot{\bf r}_j}=\sum\limits_{l=0}^\infty\sum\limits_{m=-l}^l4\pi(-i)^lj_l(kr_j)Y_{lm}^*(\Omega_{\bf k})Y_{lm}(\Omega_{{\bf r}_j}),
\end{eqnarray}
when taking into account the contribution of the spatial wave functions of the emitted photon, the initial hadron, and the final hadron. Here, $j_l(x)$ and $Y_{l m}(\Omega_{\bf x})$ are the spherical Bessel function and the spherical harmonic function, respectively.

Within the constituent quark model, the effective masses of the quarks are the crucial input parameters that affect the reliability of the results obtained for the electromagnetic properties of the isoscalar $\Xi_{c c} D^{(*)}$, $\Xi_{cc}D_{1}$ and $\Xi_{cc}D_{2}^{*}$ triple-charm molecular-type pentaquark candidates. In this paper, we adopt $m_u=0.336~{\rm GeV}$, $m_d=0.336~{\rm GeV}$, and $m_c=1.680~{\rm GeV}$ \cite{Kumar:2005ei,Li:2021ryu,Zhou:2022gra,Wang:2022tib,Wang:2022nqs,Wang:2023aob,Wang:2023ael} in the realistic calculations. To validate the reliability of the above input parameters, we will first use the constituent quark model and the above input parameters to discuss the transition magnetic moments and the magnetic moments of our focused doubly-charmed baryons and charmed mesons. It is important to note that the $D_1(2420)$ meson is a mixture of the $1^{1}P_{1}$ and $1^{3}P_{1}$ states, which satisfies the following relation \cite{Godfrey:1986wj,Matsuki:2010zy,Barnes:2002mu,Song:2015fha}
	\begin{eqnarray}
		\begin{pmatrix} \left|D_1(2430)\right\rangle  \\   \left|D_1(2420)\right\rangle \end{pmatrix}=\begin{pmatrix} \cos \theta_{1P} & \sin \theta_{1P} \\ -\sin \theta_{1P} & \cos \theta_{1P} \end{pmatrix} \begin{pmatrix} \left|1^{1}P_{1}\right\rangle  \\ \left|1^{3}P_{1}\right\rangle \end{pmatrix},
	\end{eqnarray}
where the mixing angle $\theta_{1P}$ in the heavy quark limit is $-54.7^{\circ}$ \cite{Godfrey:1986wj,Matsuki:2010zy,Barnes:2002mu,Song:2015fha}. 

In Table \ref{TMMT}, we present the transition magnetic moments and the magnetic moments of our focused doubly-charmed baryons and charmed mesons. In addition, we compare our obtained numerical results with those of other theoretical studies. From the Table \ref{TMMT}, we find that the transition magnetic moments and the magnetic moments of our focused doubly-charmed baryons and charmed mesons are in agreement with those predicted in other theoretical studies. We therefore consider the effective masses of the quarks \cite{Kumar:2005ei,Li:2021ryu,Zhou:2022gra,Wang:2022tib,Wang:2022nqs,Wang:2023aob,Wang:2023ael} to be reliable. This allows us to provide the credible theoretical insights for the experimental investigation of the electromagnetic properties of the isoscalar $\Xi_{c c} D^{(*)}$, $\Xi_{cc}D_{1}$, and $\Xi_{cc}D_{2}^{*}$ triple-charmed molecular-type pentaquark candidates.

\renewcommand\tabcolsep{0.47cm}
\renewcommand{\arraystretch}{1.50}
\begin{table}[!htbp]
\caption{Our obtained transition magnetic moments and magnetic moments of our focused doubly-charmed baryons and charmed mesons, and comparison with them obtained from other theoretical studies. Both the transition magnetic moments and the magnetic moments of the hadrons are expressed in units of $\mu_{N}=e/2m_{p}$.}\label{TMMT}
\begin{tabular}{c|c|l}
\toprule[1.0pt]
\toprule[1.0pt]
\multicolumn{3}{c}{Transition magnetic moments}\\\midrule[1.0pt]
Processes &  \multicolumn{1}{c|}{Our work}  &  \multicolumn{1}{c}{Other studies} \\\hline			
$D^{*0} \to D^{0} \gamma$ & $2.173$ &$2.134$ \cite{Wang:2023bek},\,$2.233$ \cite{Zhou:2022gra}\\
$D^{*+} \to D^{+} \gamma$ & $-0.538$ &$-0.559$ \cite{Zhou:2022gra},\,$-0.540$ \cite{Simonis:2018rld}\\
$D_{2}^{*0} \to D_{1}^{0} \gamma$ & $1.277$ & \multicolumn{1}{c}{/}\\
$D_{2}^{*+} \to D_{1}^{+} \gamma$ & $-0.452$ &\multicolumn{1}{c}{/}\\		\midrule[1.0pt]
\multicolumn{3}{c}{Magnetic moments}\\\midrule[1.0pt]
Hadrons &  \multicolumn{1}{c|}{Our work}  &  \multicolumn{1}{c}{Other studies} \\\hline		
$\Xi_{cc}^{+}$ & $0.807$ &$0.815$ \cite{Hazra:2021lpa} ,\,$0.806$ \cite{Lichtenberg:1976fi}\\ 			
$\Xi_{cc}^{++}$&$-0.124$&$-0.124$ \cite{Lichtenberg:1976fi},\,$-0.110$ \cite{Simonis:2018rld}\\	
$D^{*0}$ & $-1.489$ &$-1.485$ \cite{Wang:2023bek},\,$-1.489$ \cite{Zhou:2022gra}\\
$D^{*+}$ & $1.303$ &$1.308$ \cite{Wang:2023bek},\,$1.303$ \cite{Zhou:2022gra}\\
$D_{1}^{0}$ & $0.001$ &\multicolumn{1}{c}{/}\\
$D_{1}^{+}$ & $0.543$ &\multicolumn{1}{c}{/}\\
$D_{2}^{*0}$ & $-2.979$ &\multicolumn{1}{c}{/}\\
$D_{2}^{*+}$ & $2.141$ &\multicolumn{1}{c}{/}\\
\bottomrule[1.0pt]
\bottomrule[1.0pt]
\end{tabular}
\end{table}

In the following, we discuss the M1 radiative decay widths of the isoscalar $\Xi_{c c} D^{(*)}$, $\Xi_{cc}D_{1}$, and $\Xi_{cc}D_{2}^{*}$ triple-charm molecular-type pentaquark candidates, where both the $S$-$D$ wave mixing effect and the coupled channel effect are taken into account. The flavor wave functions of the isoscalar $\Xi_{c c} D^{(*)}$, $\Xi_{cc}D_{1}$, and $\Xi_{cc}D_{2}^{*}$ pentaquark systems are \cite{Chen:2017jjn,Wang:2019aoc}
\begin{eqnarray}\label{mf}		
|\Xi_{c c} D^{(*)}\rangle&=&-\frac{1}{\sqrt{2}}\left(|\Xi_{c c}^{++} D^{(*) 0}\rangle+|\Xi_{c c}^{+} D^{(*)+}\rangle\right),\\
|\Xi_{cc}D_{1}\rangle&=&-\frac{1}{\sqrt{2}}\left(|\Xi_{c c}^{++} D_{1}^{0}\rangle+|\Xi_{c c}^{+} D_{1}^{+}\rangle\right),\\
|\Xi_{cc}D_{2}^{*}\rangle&=&-\frac{1}{\sqrt{2}}\left(|\Xi_{c c}^{++} D_{2}^{*0}\rangle+|\Xi_{cc}^{+}D_{2}^{*+}\rangle\right),
\end{eqnarray}
and their corresponding spin wave functions can be constructed by the coupling of the spins of the constituent hadrons. Taking into account the $S$-$D$ wave mixing effect and the coupled channel effect, the relevant channels are \cite{Chen:2017jjn,Wang:2019aoc}
\begin{eqnarray}
\renewcommand\tabcolsep{0.90cm}
\renewcommand{\arraystretch}{1.50}
\begin{array}{*{3}ccc}
\hline
~~~~~~{\rm States}~~~~~~&~~~~~~J^P~~~~~~&~~~~~~{\rm Channels}~~~~~~\\
\hline
\Xi_{cc}D&1/2^-&|{}^2\mathbb{S}_{\frac{1}{2}}\rangle\nonumber\\
\Xi_{cc}D^*&1/2^-&|{}^2\mathbb{S}_{\frac{1}{2}}/{}^4\mathbb{D}_{\frac{1}{2}}\rangle\nonumber\\
\Xi_{cc}D^*&3/2^-&|{}^4\mathbb{S}_{\frac{3}{2}}/{}^2\mathbb{D}_{\frac{3}{2}}/{}^4\mathbb{D}_{\frac{3}{2}}\rangle\nonumber\\
\Xi_{cc}D_{1}&1/2^+&|{}^2\mathbb{S}_{\frac{1}{2}}/{}^4\mathbb{D}_{\frac{1}{2}}\rangle\nonumber\\
\Xi_{cc}D_{1}&3/2^+&|{}^4\mathbb{S}_{\frac{3}{2}}/{}^2\mathbb{D}_{\frac{3}{2}}/{}^4\mathbb{D}_{\frac{3}{2}}\rangle\nonumber\\
\Xi_{cc}D_{2}^{*}&3/2^+&|{}^4\mathbb{S}_{\frac{3}{2}}/{}^4\mathbb{D}_{\frac{3}{2}}/{}^6\mathbb{D}_{\frac{3}{2}}\rangle\nonumber\\
\Xi_{cc}D_{2}^{*}&5/2^+&|{}^5\mathbb{S}_{\frac{5}{2}}/{}^4\mathbb{D}_{\frac{5}{2}}/{}^6\mathbb{D}_{\frac{5}{2}}\rangle\\
\hline
\end{array}.
\end{eqnarray}
Here, the spin $S$, the orbital angular momentum $L$, and the total angular momentum $J$ of the corresponding channel are denoted as the symbol $|^{2S+1}L_J\rangle$. In our concrete calculations, the spin-orbital wave function $|{ }^{2 S+1} L_{J}\rangle$ must be expanded by incorporating the spin wave function $\left|S, m_{S}\right\rangle$ and the orbital wave function $Y_{L m_{L}}$ when calculating the transition magnetic moment and the magnetic moment of the $D$-wave channel. This can be expressed as $\left|{ }^{2 S+1} L_{J}\right\rangle=\sum_{m_{S}, m_{L}} C_{S m_{S}, L m_{L}}^{J,M} \left|S, m_{S}\right\rangle Y_{L m_{L}}$ \cite{Wang:2022nqs,Wang:2023bek,Wang:2023aob,Wang:2023ael}. In Table \ref{Masses}, we show the hadron masses \cite{ParticleDataGroup:2022pth} used in our calculations.

\renewcommand\tabcolsep{0.25cm}
\renewcommand{\arraystretch}{1.5}
\begin{table}[!htbp]
	\centering
	\caption{A summary of the hadron masses used in our calculations. The mass of the $\Xi_{cc}^{+}$ is taken from Ref. \cite{LHCb:2021eaf}, while the masses of other hadrons are sourced from the Particle Data Group \cite{ParticleDataGroup:2022pth}, and the units of the hadron masses are GeV.}\label{Masses}
	\begin{tabular}{cccc}
		\toprule[1.0pt]
		\toprule[1.0pt]
$m_N=0.938$ &$m_{\Xi_{cc}^{+}}=3.620$&
		$m_{\Xi_{cc}^{++}}=3.622$ & $m_{D^{0}}=1.865$ \\ $m_{D^{+}}=1.870$ & $m_{D^{*0}}=2.007$&
		$m_{D^{*+}}=2.010$ & $m_{D_{1}^{0}}=2.422$ \\ $m_{D_{1}^{+}}=2.426$ & $m_{D_{2}^{*0}}=2.461$&
		$m_{D_{2}^{*+}}=2.464$\\
		\bottomrule[1.0pt]
		\bottomrule[1.0pt]
	\end{tabular}
\end{table}

Combined with the mass spectrum of the isoscalar $\Xi_{c c} D^{(*)}$, $\Xi_{cc}D_{1}$, and $\Xi_{cc}D_{2}^{*}$ triple-charm molecular-type pentaquark candidates \cite{Chen:2017jjn,Wang:2019aoc}, in the present work we analyze the M1 radiative decay widths
for the $\Xi_{cc}D^{*}(3/2^-) \nolinebreak \to \nolinebreak \Xi_{cc}D(1/2^-)\gamma$, $\Xi_{cc}D_2^{*}(3/2^+)\nolinebreak\to\nolinebreak\Xi_{cc}D_1(1/2^+)\gamma$, \nolinebreak $\Xi_{cc}D_2^{*}(3/2^+)\nolinebreak\to\nolinebreak\Xi_{cc}D_1(3/2^+)\gamma$, $\Xi_{cc}D_2^{*}(5/2^+) \nolinebreak \to \nolinebreak \nolinebreak \Xi_{cc}D_1(3/2^+)\gamma$, \nolinebreak $\Xi_{cc}D_1(3/2^+) \nolinebreak \to \nolinebreak \Xi_{cc}D_1(1/2^+)\gamma$, and $\Xi_{cc}D_2^{*}(5/2^+) \to \Xi_{cc}D_2^{*}(3/2^+)\gamma$ processes. In Table~\ref{MTT}, we present the numerical results of the transition magnetic moments and the M1 radiative decay widths between the isoscalar $\Xi_{c c} D^{(*)}$, $\Xi_{cc}D_{1}$, and $\Xi_{cc}D_{2}^{*}$ triple-charm molecular-type pentaquark candidates discussed in three different cases: (I) neglecting the contribution of the spatial wave functions of the emitted photon, the doubly-charmed baryons, the charmed mesons, and the triple-charm molecular-type pentaquarks, (II) taking into account the contribution of the spatial wave functions of the emitted photon and the triple-charm molecular-type pentaquarks, and (III) taking into account the contribution of the spatial wave functions of the emitted photon, the doubly-charmed baryons, the charmed mesons, and the triple-charm molecular-type pentaquarks. In this study, we utilize the identical binding energies for the initial and final isoscalar $\Xi_{c c} D^{(*)}$, $\Xi_{cc}D_{1}$, and $\Xi_{cc}D_{2}^{*}$ triple-charm molecular-type pentaquark candidates and take three representative binding energies $-0.5$, $-6.0$, and $-12.0$ MeV to investigate their transition magnetic moments and M1 radiative decay widths. In addition, our analysis takes into account the contribution of the $S$-$D$ wave mixing effect and the coupled channel effect, corresponding to the coupled channel scenario.

\renewcommand\tabcolsep{0.66cm}
\renewcommand{\arraystretch}{1.50}
\begin{table*}[!htbp]
\caption{The obtained transition magnetic moments and M1 radiative decay widths between the isoscalar $\Xi_{c c} D^{(*)}$, $\Xi_{cc}D_{1}$, and $\Xi_{cc}D_{2}^{*}$ triple-charm molecular-type pentaquark candidates by performing the single channel scenario and the coupled channel scenario. Here, we discuss the transition magnetic moments and the M1 radiative decay widths between our discussed isoscalar triple-charm molecular-type pentaquarks with three different cases: (I) neglecting the contribution of the spatial wave functions of the emitted photon, the doubly-charmed baryons, the charmed mesons, and the triple-charm molecular-type pentaquarks, (II) taking into account the contribution of the spatial wave functions of the emitted photon and the triple-charm molecular-type pentaquarks, and (III) taking into account the contribution of the spatial wave functions of the emitted photon, the doubly-charmed baryons, the charmed mesons, and the triple-charm molecular-type pentaquarks. Meanwhile, $\Gamma_{H \to H^{\prime}\gamma}^{\rm Max}$ is the maximum value of the M1 radiative decay width obtained by varying the binding energies of the initial and final isoscalar triple-charm molecular-type pentaquark candidates.}\label{MTT}
		\begin{tabular}{c|c|c|c}\toprule[1pt]\toprule[1pt]
			\multicolumn{4}{c}{Single channel}\\\midrule[1.0pt]
			Processes&Cases&\multicolumn{1}{c|}{$\mu_{H\to H^{\prime}}(\mu_{N})$}&\multicolumn{1}{c}{$\Gamma_{H\to H^{\prime}\gamma}({\rm keV})$}\\\hline
			\multirow{3}{*}{$\Xi_{cc}D^{*}(3/2^-)\to\Xi_{cc}D(1/2^-)\gamma$}& $\mathrm{I}$&0.684&5.269, 5.269, 5.269\\
			&$\mathrm{II}$&0.461, 0.643, 0.660&2.396, 4.652, 4.898\\
			&$\mathrm{III}$&0.450, 0.627, 0.644&2.282, 4.430, 4.666\\\hline			
			\multirow{3}{*}{$\Xi_{cc}D_2^{*}(3/2^+)\to \Xi_{cc}D_1(1/2^+)\gamma$}&$\mathrm{I}$&0.439&0.0441, 0.0441, 0.0441\\
			&$\mathrm{II}$&0.422, 0.437, 0.437&0.0408, 0.0436, 0.0438\\
			&$\mathrm{III}$&0.419, 0.433, 0.434&0.0401, 0.0429, 0.0430\\\hline
			\multirow{3}{*}{$\Xi_{cc}D_2^{*}(3/2^+)\to \Xi_{cc}D_1(3/2^+)\gamma$}&$\mathrm{I}$& 0.186&0.00881, 0.00881, 0.00881\\
			&$\mathrm{II}$&0.178, 0.182, 0.181&0.00802, 0.00842, 0.00833\\
			&$\mathrm{III}$& 0.176, 0.180, 0.179&0.00789, 0.00828, 0.00819 \\	\hline	
			\multirow{3}{*}{$\Xi_{cc}D_2^{*}(5/2^+)\to \Xi_{cc}D_1(3/2^+)\gamma$} &$\mathrm{I}$& 0.372&0.0529, 0.0529, 0.0529\\
			&$\mathrm{II}$&0.356, 0.370, 0.371&0.0485, 0.0522, 0.0524\\
			&$\mathrm{III}$& 0.353, 0.367, 0.368& 0.0477, 0.0513, 0.0515\\\midrule[1.0pt]	
			Processes&Cases&\multicolumn{1}{c|}{$\mu_{H\to H^{\prime}}(\mu_{N})$}&\multicolumn{1}{c}{$\Gamma^{\rm max}_{H\to H^{\prime}\gamma}({\rm keV})$}\\\hline			
			\multirow{1}{*}{$\Xi_{cc}D_1(3/2^+)\to \Xi_{cc}D_1(1/2^+)\gamma$}&$\mathrm{I}$& 0.197& 0.0003\\	\hline	
			\multirow{1}{*}{$\Xi_{cc}D_2^{*}(5/2^+)\to \Xi_{cc}D_2^{*}(3/2^+)\gamma$} &$\mathrm{I}$& 0.360& 0.002\\\midrule[1.0pt]	
			\multicolumn{4}{c}{Coupled channel}\\\midrule[1.0pt]
			Process&Cases&\multicolumn{1}{c|}{$\mu_{H\to H^{\prime}}(\mu_{N})$}&\multicolumn{1}{c}{$\Gamma_{H\to H^{\prime}\gamma}({\rm keV})$}\\\hline
			\multirow{2}{*}{$\Xi_{cc}D^{*}  (3/2^-) \to \Xi_{cc}D/\Xi_{cc}D^{*} (1/2^-)$}&$\mathrm{II}$& $0.680,\ -0.674,\ -0.670$&5.127, 4.940, 4.888\\
			&$\mathrm{III}$& $0.447,\ -0.615,\ -0.629$&2.196, 4.094, 4.291\\\midrule[1.0pt]			
			Process&Cases&\multicolumn{1}{c|}{$\mu_{H\to H^{\prime}}(\mu_{N})$}&\multicolumn{1}{c}{$\Gamma^{\rm max}_{H\to H^{\prime}\gamma}({\rm keV})$}\\\hline
			\multirow{2}{*}{$\Xi_{cc}D_1/\Xi_{cc}D_2^{*}  (3/2^+) \to \Xi_{cc}D_1/\Xi_{cc}D_2^{*}  (1/2^+)$}&$\mathrm{II}$& $-0.267, 0.412, 0.444$&0.002\\
			&$\mathrm{III}$&$-0.266$, 0.408, 0.439&0.002\\
			\bottomrule[1pt]\bottomrule[1pt]
		\end{tabular}
	\end{table*}

From the transition magnetic moments and the M1 radiative decay widths between the isoscalar $\Xi_{c c} D^{(*)}$, $\Xi_{cc}D_{1}$, and $\Xi_{cc}D_{2}^{*}$ triple-charm molecular-type pentaquark candidates listed in Table \ref{MTT}, we find:
\begin{itemize}
  \item The M1 radiative decay widths between the isoscalar $\Xi_{c c} D^{(*)}$, $\Xi_{cc}D_{1}$, and $\Xi_{cc}D_{2}^{*}$ triple-charm molecular-type pentaquark candidates can be considered as the effective physical observables reflecting their inner structures. For example, the M1 radiative decay widths of the $\Xi_{cc}D_2^{*}(3/2^+)\to \Xi_{cc}D_1(1/2^+)\gamma$ and $\Xi_{cc}D_2^{*}(3/2^+)\to \Xi_{cc}D_1(3/2^+)\gamma$ processes are obviously different, which implies that the spin-parity quantum numbers of the isoscalar $\Xi_{cc}D_1$ molecular states can be distinguished by studying the associated M1 radiative decay widths.
  \item The contribution of the $S$-$D$ wave mixing effect and the coupled channel effect plays the minor role for the M1 radiative decay width of the $\Xi_{cc}D^{*}(3/2^-) \to \Xi_{cc}D(1/2^-) \gamma$ process. However, for the M1 radiative decay width of the $\Xi_{cc}D_1(3/2^+) \to \Xi_{cc}D_1(1/2^+) \gamma$ process, the contribution of the $S$-$D$ wave mixing effect and the coupled channel effect plays a major role, which depend on the spatial wave functions of the mixing channels.
  \item The spatial wave functions of the emitted photon, the doubly-charmed baryons, the charmed mesons, and the triple-charm molecular-type pentaquarks do not seem to have a significant effect on the transition magnetic moments and the M1 radiative decay widths between the isoscalar $\Xi_{c c} D^{(*)}$, $\Xi_{cc}D_{1}$, and $\Xi_{cc}D_{2}^{*}$ triple-charm molecular-type pentaquark candidates, especially for the contribution of the spatial wave functions of the doubly-charmed baryons and the charmed mesons.
 \item When assuming the existence of the same binding energies for the initial and final isoscalar triple-charm molecular-type pentaquarks, the phase spaces are zero for the $\Xi_{cc}D_1(3/2^+)\to \Xi_{cc}D_1(1/2^+)\gamma$ and $\Xi_{cc}D_2^{*}(5/2^+)\to \Xi_{cc}D_2^{*}(3/2^+)\gamma$ processes. In this study, we further examine their M1 radiative decay widths as the binding energies of the initial and final isoscalar triple-charm molecular-type pentaquarks vary. However, their M1 radiative decay widths are greatly suppressed, which is attributed to the small phase spaces for these processes. 
\end{itemize}

\section{The magnetic moments of the triple-charm molecular-type pentaquark}\label{sec3}					

In the previous section, we studied the M1 radiative decay widths between the isoscalar $\Xi_{c c} D^{(*)}$, $\Xi_{cc}D_{1}$, and $\Xi_{cc}D_{2}^{*}$ triple-charm molecular-type pentaquark candidates, which can be considered as the effective physical observables reflecting their inner structures. In this section, we discuss the magnetic moments of the isoscalar $\Xi_{c c} D^{(*)}$, $\Xi_{cc}D_{1}$, and $\Xi_{cc}D_{2}^{*}$ triple-charm molecular-type pentaquark candidates.

In Table~\ref{MT}, we present the numerical results of the magnetic moments of the isoscalar $\Xi_{c c} D^{(*)}$, $\Xi_{cc}D_{1}$, and $\Xi_{cc}D_{2}^{*}$ triple-charm molecular-type pentaquark candidates. Here, we investigate the contribution of the $S$-$D$ wave mixing effect and the coupled channel effect corresponding to the coupled channel scenario. For the coupled channel scenario, we consider three typical binding energies $-0.5$, $-6.0$, and $-12.0$ MeV for the isoscalar $\Xi_{c c} D^{(*)}$, $\Xi_{cc}D_{1}$, and $\Xi_{cc}D_{2}^{*}$ triple-charm molecular-type pentaquark candidates to discuss their magnetic moments.

\renewcommand\tabcolsep{0.83cm}
\renewcommand{\arraystretch}{1.50}
\begin{table}[!htbp]
\caption{The obtained magnetic moments of the isoscalar $\Xi_{c c} D^{(*)}$, $\Xi_{cc}D_{1}$, and $\Xi_{cc}D_{2}^{*}$ triple-charm molecular-type pentaquark candidates by performing the single channel scenario and the coupled channel scenario.}\label{MT}
\begin{tabular}{c|c}\toprule[1pt]\toprule[1pt]
			Molecules&\multicolumn{1}{c}{$\mu_{H}(\mu_{N})$}\\\midrule[1.0pt]
			\multicolumn{2}{c}{Single channel}\\\hline
			$\Xi_{cc}D(1/2^{-})$& 0.347\\			
			$\Xi_{cc}D^{*}(3/2^{-})$& 0.259 \\
			$\Xi_{cc}D_1 (1/2^+)$ & 0.0691\\
			$\Xi_{cc}D_1 (3/2^+)$ & 0.625 \\
			$\Xi_{cc}D_2^{*} (3/2^+)$ &$-0.579$ \\
			$\Xi_{cc}D_2^{*} (5/2^+)$ & $-0.0649$ \\
			\midrule[1.0pt]	
			\multicolumn{2}{c}{Coupled channel}\\\hline
			\multirow{1}{*}{$\Xi_{cc}D/\Xi_{cc}D^{*} (1/2^-)$} &0.334,0.323,0.320\\\hline	
			\multirow{1}{*}{$\Xi_{cc}D^{*} (3/2^-)$}& 0.249,0.249 ,0.248\\\hline	
			\multirow{1}{*}{$\Xi_{cc}D_1/\Xi_{cc}D_2^{*}  (1/2^+)$}& 0.0680,0.0681,0.0680\\\hline
			\multirow{1}{*}{$\Xi_{cc}D_{1}/\Xi_{cc}D_{2}^{*} (3/2^+)$}&0.507,$-0.0137$,$-0.183$\\	
			\bottomrule[1pt]\bottomrule[1pt]
		\end{tabular}
\end{table}		
	
As shown in Table~\ref{MT}, the magnetic moments of the isoscalar $\Xi_{c c} D^{(*)}$, $\Xi_{cc}D_{1}$, and $\Xi_{cc}D_{2}^{*}$ triple-charm molecular-type pentaquark candidates can be considered as the important aspect reflecting their inner structures, which can be used to distinguish their spin-parity quantum numbers. For example, the magnetic moments of the $\Xi_{cc}D_{1}$ state with $I(J^P)=0(1/2^+)$ and the $\Xi_{cc}D_{1}$ state with $I(J^P)=0(3/2^+)$ are obviously different, and the $\Xi_{cc}D_{2}^{*}$ state with $I(J^P)=0(3/2^+)$ and the $\Xi_{cc}D_{2}^{*}$ state with $I(J^P)=0(5/2^+)$ have different magnetic moments. Similarly to the case of the transition magnetic moments and the M1 radiative decay widths between the isoscalar $\Xi_{c c} D^{(*)}$, $\Xi_{cc}D_{1}$, and $\Xi_{cc}D_{2}^{*}$ triple-charm molecular-type pentaquark candidates, the $S$-$D$ wave mixing effect and the coupled channel effect play the minor role for the magnetic moments of several discussed isoscalar $\Xi_{c c} D^{(*)}$, $\Xi_{cc}D_{1}$, and $\Xi_{cc}D_{2}^{*}$ triple-charm molecular-type pentaquark candidates, such as the $\Xi_{cc}D$ state with $I(J^P)=0(1/2^-)$, the $\Xi_{cc}D^{*}$ state with $I(J^P)=0(3/2^-)$, and the $\Xi_{cc}D_{1}$ state with $I(J^P)=0(1/2^+)$. However, the $S$-$D$ wave mixing effect and the coupled channel effect play the important role to mediate the magnetic moment of the $\Xi_{cc}D_{1}$ state with $I(J^P)=0(3/2^+)$, and the corresponding magnetic moment is sensitive to the binding energy.

Except for the constituent quark model, other models and approaches are often used to discuss the electromagnetic properties of the hadrons in the past decades. Specially, the chiral perturbation theory is a popular way to study the electromagnetic properties of the heavy hadrons \cite{Cho:1992nt,Cheng:1992xi,Amundson:1992yp,Casalbuoni:1996pg,Jiang:2009jn,Jiang:2015xqa,Li:2017cfz,Li:2017pxa,Meng:2017dni,Kim:2018nqf,HillerBlin:2018gjw,Wang:2018gpl,Meng:2018gan,Liu:2018euh,Wang:2018cre,Shi:2018rhk,Wang:2019mhm,Li:2020uok,Shi:2021kmm,Li:2024jlq}. Thus, we hope that the electromagnetic properties of the isoscalar $\Xi_{c c} D^{(*)}$, $\Xi_{cc}D_{1}$, and $\Xi_{cc}D_{2}^{*}$ triple-charm molecular-type pentaquark candidates can be discussed by other models and approaches in the future, which can make our knowledge of the electromagnetic properties of the isoscalar $\Xi_{c c} D^{(*)}$, $\Xi_{cc}D_{1}$, and $\Xi_{cc}D_{2}^{*}$ triple-charm molecular-type pentaquark candidates become more abundant.

\section{Discussion and conclusion}\label{sec4}

In recent decades, the study of the molecular-type multiquark states has become an influential and attractive area of research in hadron physics. Currently, more predictions around the molecular-type pentaquarks have been made. Inspired by the discovery of the doubly charmed baryon $\Xi_{cc}(3620)^{++}$ by LHCb \cite{LHCb:2017iph}, the mass spectrum of the isoscalar $\Xi_{c c} D^{(*)}$, $\Xi_{cc}D_{1}$, and $\Xi_{cc}D_{2}^{*}$ triple-charmed molecular-type pentaquark candidates has been predicted in Refs. \cite{Chen:2017jjn,Wang:2019aoc}. At present, our knowledge of their properties is still not enough, and further suggestions for the experimental search for the isoscalar $\Xi_{c c} D^{(*)}$, $\Xi_{cc}D_{1}$ and $\Xi_{cc}D_{2}^{*}$ triple-charm molecular-type pentaquark candidates should be discussed.

In this work, we systematically study the electromagnetic properties including the M1 radiative decay widths and the magnetic moments of the isoscalar $\Xi_{c c} D^{(*)}$, $\Xi_{cc}D_{1}$, and $\Xi_{cc}D_{2}^{*}$ triple-charm molecular-type pentaquark candidates within the framework of the constituent quark model, where both the $S$-$D$ wave mixing effect and the coupled channel effect are taken into account. From our obtained numerical results, we conclude that the M1 radiative decay widths and the magnetic moments of the isoscalar $\Xi_{c c} D^{(*)}$, $\Xi_{cc}D_{1}$, and $\Xi_{cc}D_{2}^{*}$ triple-charm molecular-type pentaquark candidates can reflect their inner structures, which can be used to distinguish their spin-parity quantum numbers. Thus, the study of the electromagnetic properties is the important step to construct the family of the triple-charm molecular-type pentaquarks.

\begin{figure}[htbp]
  \includegraphics[width=0.45\textwidth]{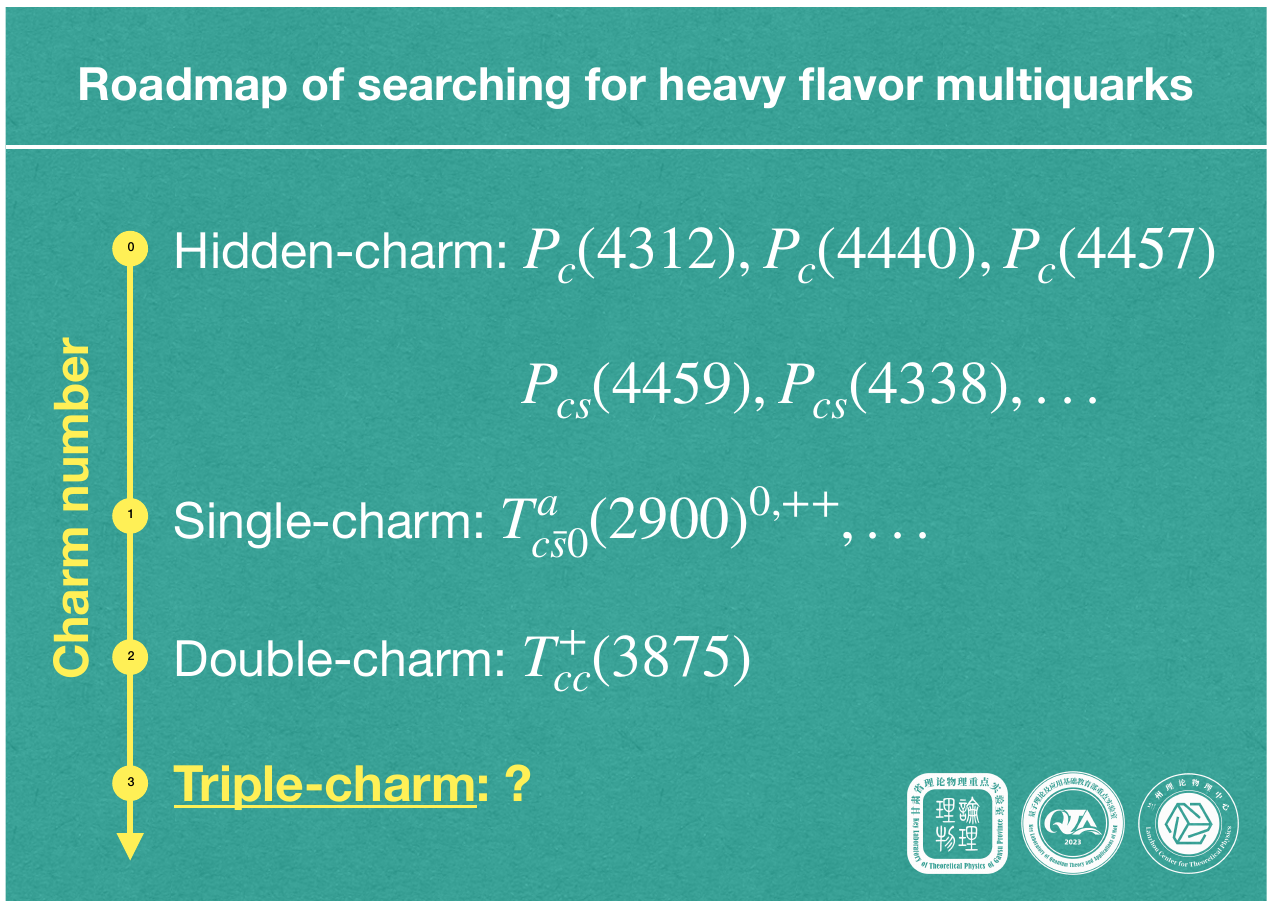}
  \caption{Roadmap of searching for heavy flavor multiquarks \cite{Liu:2013waa,Hosaka:2016pey,Chen:2016qju,Richard:2016eis,Lebed:2016hpi,Brambilla:2019esw,Liu:2019zoy,Chen:2022asf,Olsen:2017bmm,Guo:2017jvc,Meng:2022ozq}.}\label{roadmap}
\end{figure}

As emphasized in Fig. \ref{roadmap}, the single-charm, hidden-charm, and double-charm multiquark candidates have been reported in experiments  \cite{Liu:2013waa,Hosaka:2016pey,Chen:2016qju,Richard:2016eis,Lebed:2016hpi,Brambilla:2019esw,Liu:2019zoy,Chen:2022asf,Olsen:2017bmm,Guo:2017jvc,Meng:2022ozq}. So far, there have been no experimental signals of the triple-charm multiquark candidates \cite{Liu:2013waa,Hosaka:2016pey,Chen:2016qju,Richard:2016eis,Lebed:2016hpi,Brambilla:2019esw,Liu:2019zoy,Chen:2022asf,Olsen:2017bmm,Guo:2017jvc,Meng:2022ozq}. In future experiment, searching for the triple-charm multiquark candidates is an interesting and important research topic of hadron physics, and the triple-charm pentaquark state is the simplest quark configuration among the triple-charm multiquark states. As a potential experimental research task, LHCb has the opportunity to investigate the isoscalar $\Xi_{c c} D^{(*)}$, $\Xi_{cc}D_{1}$, and $\Xi_{cc}D_{2}^{*}$ triple-charm molecular-type pentaquark candidates with the accumulation of the experimental data during the high-luminosity phase of LHC \cite{Bediaga:2018lhg}. Obviously, the present work combined with the corresponding mass spectrum information \cite{Chen:2017jjn,Wang:2019aoc} can provide more comprehensive spectroscopic information when searching for the isoscalar $\Xi_{c c} D^{(*)}$, $\Xi_{cc}D_{1}$, and $\Xi_{cc}D_{2}^{*}$ triple-charm molecular-type pentaquark candidates. In addition, the theoretical study of other properties of the isoscalar $\Xi_{c c} D^{(*)}$, $\Xi_{cc}D_{1}$, and $\Xi_{cc}D_{2}^{*}$ triple-charm molecular-type pentaquark candidates is also encouraged.

\section*{ACKNOWLEDGMENTS}

This work is supported by the National Natural Science Foundation of China under Grant Nos. 12335001, 12247155 and 12247101, the China National Funds for Distinguished Young Scientists under Grant No. 11825503, National Key Research and Development Program of China under Contract No. 2020YFA0406400, the 111 Project under Grant No. B20063, the fundamental Research Funds for the Central Universities, and the project for top-notch innovative talents of Gansu province. F.L.W. is also supported by the China Postdoctoral Science Foundation under Grant No. 2022M721440.


\begin{thebibliography}{99}
		
    \bibitem{Liu:2013waa}
      X.~Liu,
      An overview of $XYZ$ new particles,
      \href{http://dx.doi.org/10.1007/s11434-014-0407-2}{Chin.\ Sci.\ Bull.\  {\bf 59}, 3815 (2014)}.

\bibitem{Hosaka:2016pey}
  A.~Hosaka, T.~Iijima, K.~Miyabayashi, Y.~Sakai, and S.~Yasui,
  Exotic hadrons with heavy flavors: $X$, $Y$, $Z$, and related states,
  \href{http://dx.doi.org/10.1093/ptep/ptw045}{Prog. Theor. Exp. Phys. {\bf 2016}, 062C01 (2016)}.

\bibitem{Chen:2016qju}
  H.~X.~Chen, W.~Chen, X.~Liu, and S.~L.~Zhu,
  The hidden-charm pentaquark and tetraquark states,
  \href{http://linkinghub.elsevier.com/retrieve/pii/S037015731630103X}{Phys.\ Rep.\  {\bf 639}, 1 (2016)}.


\bibitem{Richard:2016eis}
J.~M.~Richard,
Exotic hadrons: review and perspectives,
\href{https://link.springer.com/article/10.1007/s00601-016-1159-0}{Few Body Syst. \textbf{57}, 1185-1212 (2016)}.

\bibitem{Lebed:2016hpi}
R.~F.~Lebed, R.~E.~Mitchell and E.~S.~Swanson,
Heavy-Quark QCD Exotica,
\href{https://www.sciencedirect.com/science/article/pii/S0146641016300734?via\%3Dihub}{Prog. Part. Nucl. Phys. \textbf{93}, 143-194 (2017)}.

\bibitem{Olsen:2017bmm}
  S.~L.~Olsen, T.~Skwarnicki, and D.~Zieminska,
  Nonstandard heavy mesons and baryons: Experimental evidence,
  \href{https://journals.aps.org/rmp/abstract/10.1103/RevModPhys.90.015003}{Rev.\ Mod.\ Phys.\  {\bf 90}, 015003 (2018)}.

\bibitem{Guo:2017jvc}
  F.~K.~Guo, C.~Hanhart, U.~G.~Mei$\ss$ner, Q.~Wang, Q.~Zhao, and B.~S.~Zou,
  Hadronic molecules,
  \href{https://journals.aps.org/rmp/abstract/10.1103/RevModPhys.90.015004}{Rev.\ Mod.\ Phys.\  {\bf 90}, 015004 (2018)}.

    \bibitem{Liu:2019zoy}
      Y.~R.~Liu, H.~X.~Chen, W.~Chen, X.~Liu, and S.~L.~Zhu,
      Pentaquark and tetraquark states,
      \href{https://www.sciencedirect.com/science/article/pii/S0146641019300304?via\%3Dihub}{Prog.\ Part.\ Nucl.\ Phys.\  {\bf 107}, 237 (2019)}.

\bibitem{Brambilla:2019esw}
N.~Brambilla, S.~Eidelman, C.~Hanhart, A.~Nefediev, C.~P.~Shen, C.~E.~Thomas, A.~Vairo, and C.~Z.~Yuan,
The $XYZ$ states: Experimental and theoretical status and perspectives,
\href{https://www.sciencedirect.com/science/article/pii/S0370157320301915?via\%3Dihub}{Phys. Rep. \textbf{873}, 1 (2020)}.

\bibitem{Meng:2022ozq}
L.~Meng, B.~Wang, G.~J.~Wang and S.~L.~Zhu,
Chiral perturbation theory for heavy hadrons and chiral effective field theory for heavy hadronic molecules,
\href{https://www.sciencedirect.com/science/article/pii/S0370157323001679?via\%3Dihub}{Phys. Rept. \textbf{1019}, 1-149 (2023)}.

\bibitem{Chen:2022asf}
H.~X.~Chen, W.~Chen, X.~Liu, Y.~R.~Liu and S.~L.~Zhu,
An updated review of the new hadron states,
\href{https://iopscience.iop.org/article/10.1088/1361-6633/aca3b6}{Rept. Prog. Phys. \textbf{86}, no.2, 026201 (2023)}.
\bibitem{Wu:2010jy}
  J.~J.~Wu, R.~Molina, E.~Oset and B.~S.~Zou,
  Prediction of narrow $N^*$ and $\Lambda^*$ resonances with hidden charm above 4 GeV,
  \href{https://journals.aps.org/prl/abstract/10.1103/PhysRevLett.105.232001}{Phys.\ Rev.\ Lett.\  {\bf 105}, 232001 (2010)}.

\bibitem{Wang:2011rga}
  W.~L.~Wang, F.~Huang, Z.~Y.~Zhang, and B.~S.~Zou,
  $\Sigma_c \bar{D}$ and $\Lambda_c \bar{D}$ states in a chiral quark model,
  \href{https://journals.aps.org/prc/abstract/10.1103/PhysRevC.84.015203}{Phys.\ Rev.\ C {\bf 84}, 015203 (2011)}.

\bibitem{Yang:2011wz}
  Z.~C.~Yang, Z.~F.~Sun, J.~He, X.~Liu, and S.~L.~Zhu,
  The possible hidden-charm molecular baryons composed of anti-charmed meson and charmed baryon,
  \href{https://iopscience.iop.org/article/10.1088/1674-1137/36/1/002/meta}{Chin.\ Phys.\ C {\bf 36}, 6 (2012)}.

\bibitem{Wu:2012md}
  J.~J.~Wu, T.-S.~H.~Lee, and B.~S.~Zou,
  Nucleon resonances with hidden charm in coupled-channel Models,
  \href{https://journals.aps.org/prc/abstract/10.1103/PhysRevC.85.044002}{Phys.\ Rev.\ C {\bf 85}, 044002 (2012)}.

\bibitem{Li:2014gra}
  X.~Q.~Li and X.~Liu,
  A possible global group structure for exotic states,
  \href{https://link.springer.com/article/10.1140\%2Fepjc\%2Fs10052-014-3198-3}{Eur.\ Phys.\ J.\ C {\bf 74}, 3198 (2014)}.

\bibitem{Chen:2015loa}
  R.~Chen, X.~Liu, X.~Q.~Li, and S.~L.~Zhu,
  Identifying Exotic Hidden-Charm Pentaquarks,
  \href{https://journals.aps.org/prl/abstract/10.1103/PhysRevLett.115.132002}{Phys.\ Rev.\ Lett.\  {\bf 115}, 132002 (2015)}.

\bibitem{Karliner:2015ina}
  M.~Karliner and J.~L.~Rosner,
  New Exotic Meson and Baryon Resonances from Doubly-Heavy Hadronic Molecules,
  \href{https://journals.aps.org/prl/abstract/10.1103/PhysRevLett.115.122001}{Phys.\ Rev.\ Lett.\  {\bf 115}, 122001 (2015)}.
\bibitem{Aaij:2015tga}
  R.~Aaij {\it et al.} (LHCb Collaboration),
 Observation of $J/\psi p$ resonances consistent with pentaquark states in $\Lambda_b^0 \to J/\psi K^- p$ decays,
 \href{https://journals.aps.org/prl/abstract/10.1103/PhysRevLett.115.072001}{Phys.\ Rev.\ Lett.\  {\bf 115}, 072001 (2015)}.

\bibitem{Aaij:2019vzc}
  R.~Aaij {\it et al.} (LHCb Collaboration),
  Observation of a narrow pentaquark state, $P_c(4312)^+$, and of two-peak structure of the $P_c(4450)^+$,
 \href{https://journals.aps.org/prl/abstract/10.1103/PhysRevLett.122.222001}{Phys.\ Rev.\ Lett.\  {\bf 122}, 222001 (2019)}.

\bibitem{LHCb:2020jpq}
R.~Aaij \textit{et al.}(LHCb Collaboration),
Evidence of a $J/\psi\Lambda$ structure and observation of excited $\Xi^-$ states in the $\Xi^-_b \to J/\psi\Lambda K^-$ decay,
\href{https://www.sciencedirect.com/science/article/pii/S2095927321001717?via\%3Dihub}{Sci. Bull. \textbf{66}, 1278-1287 (2021)}.

\bibitem{LHCb:2022ogu}
R.~Aaij \textit{et al.}(LHCb Collaboration),
Observation of a $J/\psi\Lambda$ resonance consistent with a strange pentaquark candidate in $B^-\to J/\psi\Lambda\bar{p}$ decays,
\href{https://journals.aps.org/prl/abstract/10.1103/PhysRevLett.131.031901}{Phys. Rev. Lett. \textbf{131}, 031901 (2023)}.

\bibitem{Chen:2017jjn}
R.~Chen, A.~Hosaka and X.~Liu,
Prediction of triple-charm molecular pentaquarks,
\href{https://journals.aps.org/prd/abstract/10.1103/PhysRevD.96.114030}{Phys. Rev. D \textbf{96}, no.11, 114030 (2017)}.

\bibitem{Wang:2019aoc}
F.~L.~Wang, R.~Chen, Z.~W.~Liu and X.~Liu,
Possible triple-charm molecular pentaquarks from $\Xi_{cc}D_1/\Xi_{cc}D_2^*$ interactions,
\href{https://journals.aps.org/prd/abstract/10.1103/PhysRevD.99.054021}{Phys. Rev. D \textbf{99}, no.5, 054021 (2019)}.

\bibitem{LHCb:2017iph}
R.~Aaij \textit{et al.} (LHCb Collaboration),
Observation of the doubly charmed baryon $\Xi_{cc}^{++}$,
\href{https://journals.aps.org/prl/abstract/10.1103/PhysRevLett.119.112001}{Phys. Rev. Lett. \textbf{119}, no.11, 112001 (2017)}.

\bibitem{Wang:2016dzu}
G.~J.~Wang, R.~Chen, L.~Ma, X.~Liu and S.~L.~Zhu,
Magnetic moments of the hidden-charm pentaquark states,
\href{https://journals.aps.org/prd/abstract/10.1103/PhysRevD.94.094018}{Phys. Rev. D \textbf{94}, no.9, 094018 (2016)}.

\bibitem{Li:2021ryu}
M.~W.~Li, Z.~W.~Liu, Z.~F.~Sun and R.~Chen,
Magnetic moments and transition magnetic moments of $P_c$ and $P_{cs}$ states,
\href{https://journals.aps.org/prd/abstract/10.1103/PhysRevD.104.054016}{Phys. Rev. D \textbf{104}, no.5, 054016 (2021)}.

\bibitem{Ozdem:2021ugy}
U.~\"Ozdem,
Magnetic dipole moments of the hidden-charm pentaquark states: $P_c(4440)$, $P_c(4457)$ and $P_{cs}(4459)$,
\href{https://link.springer.com/article/10.1140/epjc/s10052-021-09070-3}{Eur. Phys. J. C \textbf{81}, no.4, 277 (2021)}.

\bibitem{Xu:2020flp}
Y.~J.~Xu, Y.~L.~Liu and M.~Q.~Huang,
The magnetic moment of $P_{c}(4312)$ as a $\bar{D}\Sigma _{c}$ molecular state,
\href{https://link.springer.com/article/10.1140/epjc/s10052-021-09211-8}{Eur. Phys. J. C \textbf{81}, no.5, 421 (2021)}.

\bibitem{Zhou:2022gra}
H.~Y.~Zhou, F.~L.~Wang, Z.~W.~Liu and X.~Liu,
Probing the electromagnetic properties of the $\Sigma_c^{(*)}D^{(*)}$-type doubly charmed molecular pentaquarks,
\href{https://journals.aps.org/prd/abstract/10.1103/PhysRevD.106.034034}{Phys. Rev. D \textbf{106}, no.3, 034034 (2022)}.

\bibitem{Gao:2021hmv}
F.~Gao and H.~S.~Li,
Magnetic moments of hidden-charm strange pentaquark states,
\href{https://iopscience.iop.org/article/10.1088/1674-1137/ac8651}{Chin. Phys. C \textbf{46}, no.12, 123111 (2022)}.

\bibitem{Wang:2022nqs}
F.~L.~Wang, S.~Q.~Luo, H.~Y.~Zhou, Z.~W.~Liu and X.~Liu,
Exploring the electromagnetic properties of the $\Xi_c^{(\prime,\,*)} \bar D_s^*$ and $\Omega_c^{(*)} \bar D_s^*$ molecular states,
\href{https://journals.aps.org/prd/abstract/10.1103/PhysRevD.108.034006}{Phys. Rev. D \textbf{108}, 034006 (2023)}.

\bibitem{Ozdem:2023htj}
U.~Ozdem,
Electromagnetic properties of $\bar D^{(*)}\Xi^{\prime}_c$, $\bar D^{(*)}\Lambda_c$, $\bar D_s^{(*)}\Lambda_c$ and $\bar D_s^{(*)}\Xi_c$ pentaquarks,
\href{https://www.sciencedirect.com/science/article/pii/S0370269323006019?via\%3Dihub}{Phys. Lett. B \textbf{846}, 138267 (2023)}.

\bibitem{Ozdem:2022kei}
U.~\"Ozdem,
Investigation of magnetic moment of $P_{cs}(4338)$ and $P_{cs}(4459)$ pentaquark states,
\href{https://www.sciencedirect.com/science/article/pii/S0370269323000497?via\%3Dihub}{Phys. Lett. B \textbf{836}, 137635 (2023)}.

\bibitem{Wang:2022tib}
F.~L.~Wang, H.~Y.~Zhou, Z.~W.~Liu and X.~Liu,
What can we learn from the electromagnetic properties of hidden-charm molecular pentaquarks with single strangeness?,
\href{https://journals.aps.org/prd/abstract/10.1103/PhysRevD.106.054020}{Phys. Rev. D \textbf{106}, no.5, 054020 (2022)}.

\bibitem{Wang:2023aob}
F.~L.~Wang and X.~Liu,
New type of doubly charmed molecular pentaquarks containing most strange quarks: Mass spectra, radiative decays, and magnetic moments,
\href{https://journals.aps.org/prd/abstract/10.1103/PhysRevD.108.074022}{Phys. Rev. D \textbf{108}, no.7, 074022 (2023)}.

\bibitem{Wang:2023ael}
F.~L.~Wang and X.~Liu,
Surveying the mass spectra and the electromagnetic properties of the $\Xi_c^{(\prime,\,*)}  D^*$ molecular pentaquarks,
\href{https://journals.aps.org/prd/abstract/10.1103/PhysRevD.109.014043}{Phys. Rev. D \textbf{109}, no.1, 014043 (2024)}.

\bibitem{Guo:2023fih}
F.~Guo and H.~S.~Li,
Analysis of the hidden-charm pentaquark states $P^{N^{0}}_{\psi}$ based on magnetic moment and transition magnetic moment,
\href{https://arxiv.org/abs/2304.10981}{arXiv:2304.10981}.

\bibitem{Li:2024wxr}
H.~S.~Li, F.~Guo, Y.~D.~Lei and F.~Gao,
Magnetic moments and axial charges of the octet hidden-charm molecular pentaquark family,
\href{https://arxiv.org/abs/2401.14767}{arXiv:2401.14767}.

\bibitem{Li:2024jlq}
H.~S.~Li,
Ab initio calculation of molecular pentaquark magnetic moments in heavy pentaquark chiral perturbation theory,
\href{https://arxiv.org/abs/2401.14759}{arXiv:2401.14759}.

\bibitem{Ozdem:2024jty}
U.~\"Ozdem,
Analysis of the isospin eigenstate $\bar D \Sigma_c$, $\bar D^{*} \Sigma_c$, and $\bar D \Sigma_c^{*}$ pentaquarks by their electromagnetic properties,
\href{https://arxiv.org/abs/2401.12678}{arXiv:2401.12678}.

\bibitem{Ozdem:2024rqx}
U.~\"Ozdem,
Elucidating the nature of hidden-charm pentaquark states with spin-32 through their electromagnetic form factors,
\href{https://www.sciencedirect.com/science/article/pii/S0370269324001096?via\%3Dihub}{Phys. Lett. B \textbf{851} (2024), 138551}.

\bibitem{Bediaga:2018lhg}
R.~Aaij \textit{et al.} (LHCb Collaboration),
Physics case for an LHCb Upgrade II-Opportunities in flavour physics, and beyond, in the HL-LHC era,
\href{https://arxiv.org/abs/1808.08865}{arXiv:1808.08865}.

\bibitem{Tornqvist:1993ng}
  N.~A.~Tornqvist,
  From the deuteron to deusons, an analysis of deuteron-like meson-meson bound states,
   \href{https://link.springer.com/article/10.1007\%2FBF01413192}{Z.\ Phys.\ C {\bf 61}, 525 (1994)}.

\bibitem{Machleidt:1987hj}
R.~Machleidt, K.~Holinde and C.~Elster,
The Bonn Meson Exchange Model for the Nucleon Nucleon Interaction,
\href{https://www.sciencedirect.com/science/article/abs/pii/S0370157387800029?via\%3Dihub}{Phys. Rept. \textbf{149}, 1-89 (1987)}.

\bibitem{Epelbaum:2008ga}
E.~Epelbaum, H.~W.~Hammer and U.~G.~Meissner,
Modern Theory of Nuclear Forces,
\href{https://journals.aps.org/rmp/abstract/10.1103/RevModPhys.81.1773}{Rev. Mod. Phys. \textbf{81}, 1773-1825 (2009)}.

\bibitem{Esposito:2014rxa}
A.~Esposito, A.~L.~Guerrieri, F.~Piccinini, A.~Pilloni and A.~D.~Polosa,
Four-Quark Hadrons: an Updated Review,
\href{https://www.worldscientific.com/doi/abs/10.1142/S0217751X15300021}{Int. J. Mod. Phys. A \textbf{30}, 1530002 (2015)}.

\bibitem{Tornqvist:1993vu}
  N.~A.~Tornqvist,
  On deusons or deuteron-like meson-meson bound states,
  \href{https://link.springer.com/article/10.1007\%2FBF02734018}{Nuovo Cim. Soc. Ital. Fis.  {\bf 107A}, 2471 (1994)}.

\bibitem{Wang:2019nwt}
  F.~L.~Wang, R.~Chen, Z.~W.~Liu, and X.~Liu,
  Probing new types of $P_c$ states inspired by the interaction between $S-$wave charmed baryon and anti-charmed meson in a $\bar T$ doublet,
  \href{https://journals.aps.org/prc/abstract/10.1103/PhysRevC.101.025201}{Phys.\ Rev.\ C {\bf 101},  025201 (2020)}.

\bibitem{BaBar:2008flx}
B.~Aubert \textit{et al.} (BaBar Collaboration),
Evidence for $X(3872) \to \psi(2S) \gamma$ in $B^\pm \to X(3872) K^\pm$ decays, and a study of $B \to c \bar{c} \gamma K$,
\href{https://journals.aps.org/prl/abstract/10.1103/PhysRevLett.102.132001}{Phys. Rev. Lett. \textbf{102}, 132001 (2009)}.

\bibitem{LHCb:2014jvf}
R.~Aaij \textit{et al.} (LHCb Collaboration),
Evidence for the decay $X(3872)\rightarrow\psi(2S)\gamma$,
\href{https://doi.org/10.1016/j.nuclphysb.2014.06.011}{Nucl. Phys. B \textbf{886}, 665 (2014)}.

\bibitem{Belle:2011wdj}
V.~Bhardwaj \textit{et al.} (Belle Collaboration),
Observation of $X(3872)\to J/\psi \gamma$ and search for $X(3872)\to\psi'\gamma$ in B decays,
\href{https://journals.aps.org/prl/abstract/10.1103/PhysRevLett.107.091803}{Phys. Rev. Lett. \textbf{107}, 091803 (2011)}.

\bibitem{BESIII:2020nbj}
M.~Ablikim \textit{et al.} (BESIII Collaboration),
Study of open-charm decays and radiative transitions of the $X(3872)$,
\href{https://journals.aps.org/prl/abstract/10.1103/PhysRevLett.124.242001}{Phys. Rev. Lett. \textbf{124}, no.24, 242001 (2020)}.

\bibitem{Schlumpf:1993rm}
F.~Schlumpf,
Magnetic moments of the baryon decuplet in a relativistic quark model,
\href{https://journals.aps.org/prd/abstract/10.1103/PhysRevD.48.4478}{Phys. Rev. D \textbf{48}, 4478-4480 (1993)}.

\bibitem{Schlumpf:1992vq}
F.~Schlumpf,
Relativistic constituent quark model of electroweak properties of baryons,
\href{https://journals.aps.org/prd/abstract/10.1103/PhysRevD.47.4114}{Phys. Rev. D \textbf{47}, 4114 (1993)};
\href{https://journals.aps.org/prd/abstract/10.1103/PhysRevD.49.6246.2}{[Phys. Rev. D \textbf{49}, 6246 (1994)]}.

\bibitem{Cheng:1997kr}
T.~P.~Cheng and L.~F.~Li,
Why naive quark model can yield a good account of the baryon magnetic moments,
\href{https://journals.aps.org/prl/abstract/10.1103/PhysRevLett.80.2789}{Phys. Rev. Lett. \textbf{80}, 2789-2792 (1998)}.

\bibitem{Ha:1998gf}
P.~Ha and L.~Durand,
Baryon magnetic moments in a QCD based quark model with loop corrections,
\href{https://journals.aps.org/prd/abstract/10.1103/PhysRevD.58.093008}{Phys. Rev. D \textbf{58}, 093008 (1998)}.

\bibitem{Liu:2003ab}
Y.~R.~Liu, P.~Z.~Huang, W.~Z.~Deng, X.~L.~Chen and S.~L.~Zhu,
Pentaquark magnetic moments in different models,
\href{https://journals.aps.org/prc/abstract/10.1103/PhysRevC.69.035205}{Phys. Rev. C \textbf{69}, 035205 (2004)}.

\bibitem{Huang:2004tn}
P.~Z.~Huang, Y.~R.~Liu, W.~Z.~Deng, X.~L.~Chen and S.~L.~Zhu,
Heavy pentaquarks,
\href{https://journals.aps.org/prd/abstract/10.1103/PhysRevD.70.034003}{Phys. Rev. D \textbf{70}, 034003 (2004)}.

\bibitem{Zhu:2004xa}
S.~L.~Zhu,
Pentaquarks,
\href{https://www.worldscientific.com/doi/abs/10.1142/S0217751X04019676}{Int. J. Mod. Phys. A \textbf{19}, 3439-3469 (2004)}.

\bibitem{Kumar:2005ei}
S.~Kumar, R.~Dhir and R.~C.~Verma,
Magnetic moments of charm baryons using effective mass and screened charge of quarks,
\href{https://iopscience.iop.org/article/10.1088/0954-3899/31/2/006}{J. Phys. G \textbf{31}, 141-147 (2005)}.

\bibitem{Haghpayma:2006hu}
A.~R.~Haghpayma,
Magnetic Moment of the Pentaquark $\Theta^+$ State,
\href{https://arxiv.org/abs/hep-ph/0609253}{arXiv:hep-ph/0609253}.

\bibitem{Ramalho:2009gk}
G.~Ramalho, K.~Tsushima and F.~Gross,
A Relativistic quark model for the Omega-electromagnetic form factors,
\href{https://journals.aps.org/prd/abstract/10.1103/PhysRevD.80.033004}{Phys. Rev. D \textbf{80}, 033004 (2009)}.

\bibitem{Dhir:2009ax}
R.~Dhir and R.~C.~Verma,
Magnetic Moments of ($J^P = 3/2^+$) Heavy Baryons Using Effective Mass Scheme,
\href{https://link.springer.com/article/10.1140/epja/i2009-10872-8}{Eur. Phys. J. A \textbf{42}, 243-249 (2009)}.

\bibitem{Majethiya:2009vx}
A.~Majethiya, B.~Patel and P.~C.~Vinodkumar,
Radiative decays of single heavy flavour baryons,
\href{https://link.springer.com/article/10.1140/epja/i2009-10880-8}{Eur. Phys. J. A \textbf{42}, 213-218 (2009)}.

\bibitem{Sharma:2010vv}
N.~Sharma, H.~Dahiya, P.~K.~Chatley and M.~Gupta,
Spin $\frac{1}{2}^+$, spin $\frac{3}{2}^+$ and transition magnetic moments of low lying and charmed baryons,
\href{https://journals.aps.org/prd/abstract/10.1103/PhysRevD.81.073001}{Phys. Rev. D \textbf{81}, 073001 (2010)}.

\bibitem{Sharma:2012jqz}
N.~Sharma, A.~Martinez Torres, K.~P.~Khemchandani and H.~Dahiya,
Magnetic moments of the low-lying ${1/2}^-$ octet baryon resonances,
\href{https://link.springer.com/article/10.1140/epja/i2013-13011-2}{Eur. Phys. J. A \textbf{49}, 11 (2013)}.

\bibitem{Dhir:2013nka}
R.~Dhir, C.~S.~Kim and R.~C.~Verma,
Magnetic Moments of Bottom Baryons: Effective mass and Screened Charge,
\href{https://journals.aps.org/prd/abstract/10.1103/PhysRevD.88.094002}{Phys. Rev. D \textbf{88}, 094002 (2013)}.

\bibitem{Ghalenovi:2014swa}
Z.~Ghalenovi, A.~A.~Rajabi, S.~x.~Qin and D.~H.~Rischke,
Ground-State Masses and Magnetic Moments of Heavy Baryons,
\href{https://www.worldscientific.com/doi/abs/10.1142/S0217732314501065}{Mod. Phys. Lett. A \textbf{29}, 1450106 (2014)}.

\bibitem{Girdhar:2015gsa}
A.~Girdhar, H.~Dahiya and M.~Randhawa,
Magnetic moments of $J^P=\frac{3}{2}^+$ decuplet baryons using effective quark masses in chiral constituent quark model,
\href{https://journals.aps.org/prd/abstract/10.1103/PhysRevD.92.033012}{Phys. Rev. D \textbf{92}, 033012 (2015)}.

\bibitem{Majethiya:2011ry}
A.~Majethiya, K.~Thakkar and P.~C.~Vinodkumar,
Spectroscopy and decay properties of  $\Sigma_{b}, \Lambda_{b}$ baryons in quark-diquark model,
\href{https://www.sciencedirect.com/science/article/pii/S057790731630243X?via\%3Dihub}{Chin. J. Phys. \textbf{54}, 495-502 (2016)}.

\bibitem{Thakkar:2016sog}
K.~Thakkar, A.~Majethiya and P.~C.~Vinodkumar,
Magnetic moments of baryons containing all heavy quarks in the quark-diquark model,
\href{https://link.springer.com/article/10.1140/epjp/i2016-16339-4}{Eur. Phys. J. Plus \textbf{131}, 339 (2016)}.

\bibitem{Shah:2016nxi}
Z.~Shah, K.~Thakkar, A.~K.~Rai and P.~C.~Vinodkumar,
Mass spectra and Regge trajectories of $\Lambda_{c}^{+}$, $\Sigma_{c}^{0}$, $\Xi_{c}^{0}$ and $\Omega_{c}^{0}$ baryons,
\href{https://iopscience.iop.org/article/10.1088/1674-1137/40/12/123102}{Chin. Phys. C \textbf{40}, 123102 (2016)}.

\bibitem{Shah:2016vmd}
Z.~Shah, K.~Thakkar and A.~K.~Rai,
Excited State Mass spectra of doubly heavy baryons $\Omega_{cc}$, $\Omega_{bb}$ and $\Omega_{bc}$,
\href{https://link.springer.com/article/10.1140/epjc/s10052-016-4379-z}{Eur. Phys. J. C \textbf{76}, 530 (2016)}.

\bibitem{Kaur:2016kan}
A.~Kaur, P.~Gupta and A.~Upadhyay,
Properties of $J^{P}=1/2^{+}$ baryon octets at low energy,
\href{https://doi.org/10.1093/ptep/ptx068}{PTEP \textbf{2017}, 063B02 (2017)}.

\bibitem{Shah:2018bnr}
Z.~Shah and A.~Kumar Rai,
Spectroscopy of the $\Omega_{ccb}$ baryon in the hypercentral constituent quark model,
\href{https://iopscience.iop.org/article/10.1088/1674-1137/42/5/053101}{Chin. Phys. C \textbf{42}, 053101 (2018)}.

\bibitem{Gandhi:2018lez}
K.~Gandhi, Z.~Shah and A.~K.~Rai,
Decay properties of singly charmed baryons,
\href{https://link.springer.com/article/10.1140/epjp/i2018-12318-1}{Eur. Phys. J. Plus \textbf{133}, 512 (2018)}.

\bibitem{Dahiya:2018ahb}
H.~Dahiya,
Transition magnetic moments of $J^P=\frac{3}{2}^+$ decuplet to $J^P=\frac{1}{2}^+$ octet baryons in the chiral constituent quark model,
\href{https://iopscience.iop.org/article/10.1088/1674-1137/42/9/093102}{Chin. Phys. C \textbf{42}, 093102 (2018)}.

\bibitem{Simonis:2018rld}
V.~Simonis,
Improved predictions for magnetic moments and M1 decay widths of heavy hadrons,
\href{https://arxiv.org/abs/1803.01809}{arXiv:1803.01809}.

\bibitem{Ghalenovi:2018fxh}
Z.~Ghalenovi and M.~Moazzen Sorkhi,
Mass spectra and decay properties of $ \Sigma_{{b}}^{}$ and $ \Lambda_{{b}}^{}$ baryons in a quark model,
\href{https://link.springer.com/article/10.1140/epjp/i2018-12111-2}{Eur. Phys. J. Plus \textbf{133}, 301 (2018)}.

\bibitem{Gandhi:2019bju}
K.~Gandhi and A.~K.~Rai,
Spectrum of strange singly charmed baryons in the constituent quark model,
\href{https://www.sciencedirect.com/science/article/abs/pii/0370269394914664?via\%3Dihub}{Eur. Phys. J. Plus \textbf{135}, 213 (2020)}.

\bibitem{Rahmani:2020pol}
S.~Rahmani, H.~Hassanabadi and H.~Sobhani,
Mass and decay properties of double heavy baryons with a phenomenological potential model,
\href{https://link.springer.com/article/10.1140/epjc/s10052-020-7867-0}{Eur. Phys. J. C \textbf{80}, 312 (2020)}.

\bibitem{Hazra:2021lpa}
A.~Hazra, S.~Rakshit and R.~Dhir,
Radiative M1 transitions of heavy baryons: Effective quark mass scheme,
\href{https://journals.aps.org/prd/abstract/10.1103/PhysRevD.104.053002}{Phys. Rev. D \textbf{104}, 053002 (2021)}.

\bibitem{Menapara:2021dzi}
C.~Menapara and A.~K.~Rai,
Spectroscopic investigation of light strange $S=-1$ $\Lambda$, $\Sigma$ and $S=-2$ $\Xi$ baryons,
\href{https://iopscience.iop.org/article/10.1088/1674-1137/abf4f4}{Chin. Phys. C \textbf{45}, 063108 (2021)}.

\bibitem{Deng:2021gnb}
C.~Deng and S.~L.~Zhu,
$T_{cc}^+$ and its partners,
\href{https://journals.aps.org/prd/abstract/10.1103/PhysRevD.105.054015}{Phys. Rev. D \textbf{105}, no.5, 054015 (2022)}.

\bibitem{Menapara:2021vug}
C.~Menapara and A.~K.~Rai,
Spectroscopic Study of Strangeness$=-3$ $\Omega^{-}$ Baryon,
\href{https://iopscience.iop.org/article/10.1088/1674-1137/ac78d1}{Chin. Phys. C \textbf{46}, 103102 (2022)}.

\bibitem{Mutuk:2021epz}
H.~Mutuk,
The status of $\Xi _\mathrm{{cc}}^{++}$ baryon: investigating quark-diquark model,
\href{https://link.springer.com/article/10.1140/epjp/s13360-021-02256-4}{Eur. Phys. J. Plus \textbf{137}, 10 (2022)}.

\bibitem{Kakadiya:2022pin}
A.~Kakadiya, Z.~Shah and A.~K.~Rai,
Spectroscopy of $\Omega_{ccc}$ and $\Omega_{bbb}$ baryons,
\href{https://www.worldscientific.com/doi/10.1142/S0217751X22502256}{Int. J. Mod. Phys. A \textbf{37}, no.36, 2250225 (2022)}.

\bibitem{Menapara:2022ksj}
C.~Menapara and A.~K.~Rai,
Spectroscopy of light baryons: $\Delta$ resonances,
\href{https://www.worldscientific.com/doi/10.1142/S0217751X22501779}{Int. J. Mod. Phys. A \textbf{37}, no.27, 2250177 (2022)}.

\bibitem{Mohan:2022sxm}
B.~Mohan, T.~M.~S., A.~Hazra and R.~Dhir,
Screening of the quark charge and mixing effects on transition moments and M1 decay widths of baryons,
\href{https://journals.aps.org/prd/abstract/10.1103/PhysRevD.106.113007}{Phys. Rev. D \textbf{106}, no.11, 113007 (2022)}.

\bibitem{An:2022qpt}
H.~T.~An, S.~Q.~Luo, Z.~W.~Liu and X.~Liu,
Spectroscopy behavior of fully heavy tetraquarks,
\href{https://link.springer.com/article/10.1140/epjc/s10052-023-11847-7}{Eur. Phys. J. C \textbf{83}, 740 (2023)}.

\bibitem{Wu:2022gie}
T.~W.~Wu and Y.~L.~Ma,
Doubly heavy tetraquark multiplets as heavy antiquark-diquark symmetry partners of heavy baryons,
\href{https://journals.aps.org/prd/abstract/10.1103/PhysRevD.107.L071501}{Phys. Rev. D \textbf{107}, no.7, L071501 (2023)}.

\bibitem{Wang:2023bek}
F.~L.~Wang, S.~Q.~Luo and X.~Liu,
Radiative decays and magnetic moments of the predicted $B_c$-like molecules,
\href{https://journals.aps.org/prd/abstract/10.1103/PhysRevD.107.114017}{Phys. Rev. D \textbf{107}, no.11, 114017 (2023)}.
\bibitem{Dey:1994qi}
J.~Dey, V.~Shevchenko, P.~Volkovitsky and M.~Dey,
Radiative decays of $S$-wave charmed baryons,
\href{https://www.sciencedirect.com/science/article/abs/pii/0370269394914664?via\%3Dihub}{Phys. Lett. B \textbf{337}, 185-188 (1994)}.

\bibitem{ParticleDataGroup:2022pth}
R.~L.~Workman \textit{et al.} [Particle Data Group],
Review of Particle Physics,
\href{https://academic.oup.com/ptep/article/2022/8/083C01/6651666?login=true}{PTEP \textbf{2022}, 083C01 (2022)}.

\bibitem{Yu:2022lel}
G.~L.~Yu, Z.~Y.~Li, Z.~G.~Wang, J.~Lu and M.~Yan,
Systematic analysis of doubly charmed baryons $\Xi _{cc}$ and $\Omega _{cc}$,
\href{https://link.springer.com/article/10.1140/epja/s10050-023-01044-1}{Eur. Phys. J. A \textbf{59}, no.6, 126 (2023)}.

\bibitem{Godfrey:2015dva}
S.~Godfrey and K.~Moats,
Properties of Excited Charm and Charm-Strange Mesons,
\href{https://journals.aps.org/prd/abstract/10.1103/PhysRevD.93.034035}{Phys. Rev. D \textbf{93}, no.3, 034035 (2016)}.

\bibitem{Khersonskii:1988krb}
V.~K.~Khersonskii, A.~N.~Moskalev and D.~A.~Varshalovich,
Quantum Theory Of Angular Momentum,
\href{https://www.worldscientific.com/worldscibooks/10.1142/0270#t=aboutBook}{World Scientific Publishing Company, Singapore, 1988}.

\bibitem{Godfrey:1986wj}
S.~Godfrey and R.~Kokoski,
The Properties of p Wave Mesons with One Heavy Quark,
\href{https://journals.aps.org/prd/abstract/10.1103/PhysRevD.43.1679}{Phys. Rev. D \textbf{43}, 1679-1687 (1991)}.

\bibitem{Matsuki:2010zy}
T.~Matsuki, T.~Morii and K.~Seo,
Mixing angle between $^3P_1$ and $^1P_1$ in HQET,
\href{https://academic.oup.com/ptp/article/124/2/285/1829185}{Prog. Theor. Phys. \textbf{124}, 285-292 (2010)}.

\bibitem{Barnes:2002mu}
T.~Barnes, N.~Black and P.~R.~Page,
Strong decays of strange quarkonia,
\href{https://journals.aps.org/prd/abstract/10.1103/PhysRevD.68.054014}{Phys. Rev. D \textbf{68}, 054014 (2003)}.

\bibitem{Song:2015fha}
Q.~T.~Song, D.~Y.~Chen, X.~Liu and T.~Matsuki,
Higher radial and orbital excitations in the charmed meson family,
\href{https://journals.aps.org/prd/abstract/10.1103/PhysRevD.92.074011}{Phys. Rev. D \textbf{92}, no.7, 074011 (2015)}.

\bibitem{Lichtenberg:1976fi}
D.~B.~Lichtenberg,
Magnetic Moments of Charmed Baryons in the Quark Model,
\href{https://journals.aps.org/prd/abstract/10.1103/PhysRevD.15.345}{Phys. Rev. D \textbf{15}, 345 (1977)}.

\bibitem{LHCb:2021eaf}
R.~Aaij \textit{et al.} [LHCb],
Search for the doubly charmed baryon $ {\varXi}_{cc}^{+} $ in the $ {\varXi}_c^{+}{\pi}^{-}{\pi}^{+} $ final state,
\href{https://link.springer.com/article/10.1007/JHEP12(2021)107}{JHEP \textbf{12} (2021), 107}.

\bibitem{Cho:1992nt}
P.~L.~Cho and H.~Georgi,
Electromagnetic interactions in heavy hadron chiral theory,
\href{https://www.sciencedirect.com/science/article/abs/pii/037026939291340F?via\%3Dihub}{Phys. Lett. B \textbf{296} (1992), 408-414}.		

\bibitem{Cheng:1992xi}
H.~Y.~Cheng, C.~Y.~Cheung, G.~L.~Lin, Y.~C.~Lin, T.~M.~Yan and H.~L.~Yu,
Chiral Lagrangians for radiative decays of heavy hadrons,
\href{https://journals.aps.org/prd/abstract/10.1103/PhysRevD.47.1030}{Phys. Rev. D \textbf{47} (1993), 1030-1042}.		

\bibitem{Amundson:1992yp}
J.~F.~Amundson, C.~G.~Boyd, E.~E.~Jenkins, M.~E.~Luke, A.~V.~Manohar, J.~L.~Rosner, M.~J.~Savage and M.~B.~Wise,
Radiative D* decay using heavy quark and chiral symmetry,
\href{https://linkinghub.elsevier.com/retrieve/pii/0370269392913416}{	Phys. Lett. B \textbf{296} (1992), 415-419}.		


\bibitem{Casalbuoni:1996pg}
R.~Casalbuoni, A.~Deandrea, N.~Di Bartolomeo, R.~Gatto, F.~Feruglio and G.~Nardulli,
Phenomenology of heavy meson chiral Lagrangians,
\href{https://linkinghub.elsevier.com/retrieve/pii/S0370157396000270}{Phys. Rept. \textbf{281} (1997), 145-238}.

\bibitem{Jiang:2009jn}
F.~J.~Jiang and B.~C.~Tiburzi,
Hyperon Electromagnetic Properties in Two-Flavor Chiral Perturbation Theory,
\href{https://journals.aps.org/prd/abstract/10.1103/PhysRevD.81.034017}{Phys. Rev. D \textbf{81} (2010), 034017}.		

\bibitem{Jiang:2015xqa}
N.~Jiang, X.~L.~Chen and S.~L.~Zhu,
Electromagnetic decays of the charmed and bottom baryons in chiral perturbation theory,
\href{https://journals.aps.org/prd/abstract/10.1103/PhysRevD.92.054017}{Phys. Rev. D \textbf{92} (2015) no.5, 054017}.

\bibitem{Li:2017cfz}
H.~S.~Li, L.~Meng, Z.~W.~Liu and S.~L.~Zhu,
Magnetic moments of the doubly charmed and bottom baryons,
\href{https://journals.aps.org/prd/abstract/10.1103/PhysRevD.96.076011}{Phys. Rev. D \textbf{96} (2017) no.7, 076011}.

\bibitem{Li:2017pxa}
H.~S.~Li, L.~Meng, Z.~W.~Liu and S.~L.~Zhu,
Radiative decays of the doubly charmed baryons in chiral perturbation theory,		
\href{https://linkinghub.elsevier.com/retrieve/pii/S0370269317310080}{Phys. Lett. B \textbf{777} (2018), 169-176}.

\bibitem{Meng:2017dni}
L.~Meng, H.~S.~Li, Z.~W.~Liu and S.~L.~Zhu,
Magnetic moments of the spin-${3\over 2}$ doubly heavy baryons,		
\href{https://link.springer.com/article/10.1140/epjc/s10052-017-5447-8}{Eur. Phys. J. C \textbf{77} (2017) no.12, 869}.

\bibitem{Kim:2018nqf}
J.~Y.~Kim and H.~C.~Kim,
Electromagnetic form factors of singly heavy baryons in the self-consistent SU(3) chiral quark-soliton model,
\href{https://journals.aps.org/prd/abstract/10.1103/PhysRevD.97.114009}{Phys. Rev. D \textbf{97} (2018) no.11, 114009}.			

\bibitem{HillerBlin:2018gjw}
A.~N.~Hiller Blin, Z.~F.~Sun and M.~J.~Vicente Vacas,
Electromagnetic form factors of spin 1/2 doubly charmed baryons,
\href{https://journals.aps.org/prd/abstract/10.1103/PhysRevD.98.054025}{Phys. Rev. D \textbf{98} (2018) no.5, 054025}.	

\bibitem{Wang:2018gpl}
G.~J.~Wang, L.~Meng, H.~S.~Li, Z.~W.~Liu and S.~L.~Zhu,
Magnetic moments of the spin-$\frac{1}{2}$ singly charmed baryons in chiral perturbation theory,		
\href{https://journals.aps.org/prd/abstract/10.1103/PhysRevD.98.054026}{Phys. Rev. D \textbf{98} (2018) no.5, 054026}.

\bibitem{Meng:2018gan}
L.~Meng, G.~J.~Wang, C.~Z.~Leng, Z.~W.~Liu and S.~L.~Zhu,
Magnetic moments of the spin-${3\over 2}$ singly heavy baryons,
\href{https://journals.aps.org/prd/abstract/10.1103/PhysRevD.98.094013}{Phys. Rev. D \textbf{98} (2018) no.9, 094013}.		

\bibitem{Liu:2018euh}
M.~Z.~Liu, Y.~Xiao and L.~S.~Geng,
Magnetic moments of the spin-1/2 doubly charmed baryons in covariant baryon chiral perturbation theory,		
\href{https://journals.aps.org/prd/abstract/10.1103/PhysRevD.98.014040}{Phys. Rev. D \textbf{98} (2018) no.1, 014040}.

\bibitem{Wang:2018cre}
G.~J.~Wang, L.~Meng and S.~L.~Zhu,
Radiative decays of the singly heavy baryons in chiral perturbation theory,
\href{https://journals.aps.org/prd/abstract/10.1103/PhysRevD.99.034021}{Phys. Rev. D \textbf{99} (2019) no.3, 034021}.		

\bibitem{Shi:2018rhk}
R.~X.~Shi, Y.~Xiao and L.~S.~Geng,
Magnetic moments of the spin-1/2 singly charmed baryons in covariant baryon chiral perturbation theory,
\href{https://journals.aps.org/prd/abstract/10.1103/PhysRevD.100.054019}{Phys. Rev. D \textbf{100} (2019) no.5, 054019}.
	
\bibitem{Wang:2019mhm}
B.~Wang, B.~Yang, L.~Meng and S.~L.~Zhu,
Radiative transitions and magnetic moments of the charmed and bottom vector mesons in chiral perturbation theory,
\href{https://journals.aps.org/prd/abstract/10.1103/PhysRevD.100.016019}{Phys. Rev. D \textbf{100} (2019) no.1, 016019}.		

\bibitem{Li:2020uok}
H.~S.~Li and W.~L.~Yang,
Spin-$\frac{3}{2}$ doubly charmed baryon contribution to the magnetic moments of the spin-$\frac{1}{2}$ doubly charmed baryons,
\href{https://journals.aps.org/prd/abstract/10.1103/PhysRevD.103.056024}{Phys. Rev. D \textbf{103} (2021) no.5, 056024}.

\bibitem{Shi:2021kmm}
R.~X.~Shi and L.~S.~Geng,
Magnetic moments of the spin-$\frac{3}{2}$ doubly charmed baryons in covariant baryon chiral perturbation theory,
\href{https://journals.aps.org/prd/abstract/10.1103/PhysRevD.103.114004}{Phys. Rev. D \textbf{103} (2021) no.11, 114004}.		


\end{thebibliography}
\end{document}